\documentclass[twocolumn, pra, a4paper, 10pt, superscriptaddress, floatfix]{revtex4-1}
\usepackage{graphicx}
\usepackage{xcolor}
\usepackage{physics}
\usepackage{nicefrac}
\usepackage{bm}
\usepackage{float}
\usepackage[utf8]{inputenc}
\usepackage[colorlinks, citecolor=blue, linkcolor=black]{hyperref}
\usepackage{ulem}
\usepackage{booktabs}
\usepackage{multirow}
\usepackage{xspace}

\begin{document}
\title{A planar defect spin sensor in a two-dimensional material susceptible to strain and electric fields}

\author{P.~Udvarhelyi}
\affiliation{Wigner Research Centre for Physics, P.O.\ Box 49, H-1525 Budapest, Hungary}
\affiliation{Budapest University of Technology and Economics, Institute of Physics, Department of Atomic Physics, M\H{u}egyetem rakpart 3., 1111 Budapest, Hungary}
\author{T. Clua-Provost}
\author{A. Durand}
\affiliation{Laboratoire Charles Coulomb, Université de Montpellier and CNRS, 34095 Montpellier, France}
\author{J. Li}
\author{J.~H.~Edgar}
\affiliation{Tim Taylor Department of Chemical Engineering, Kansas State University, Kansas 66506, USA}
\author{B. Gil}
\author{G. Cassabois}
\author{V. Jacques}
\affiliation{Laboratoire Charles Coulomb, Université de Montpellier and CNRS, 34095 Montpellier, France}
\author{A. Gali}
\affiliation{Wigner Research Centre for Physics, P.O.\ Box 49, H-1525 Budapest, Hungary}
\affiliation{Budapest University of Technology and Economics, Institute of Physics, Department of Atomic Physics, M\H{u}egyetem rakpart 3., 1111 Budapest, Hungary}

\begin{abstract}
The boron-vacancy spin defect ($\text{V}_\text{B}^{-}$) in hexagonal boron nitride (hBN) has a great potential as a quantum sensor in a two-dimensional material that can directly probe various external perturbations in atomic-scale proximity to the quantum sensing layer.
Here, we apply first principles calculations to determine the coupling of the $\text{V}_\text{B}^{-}$ electronic spin to strain and electric fields. Our work unravels the interplay between local piezoelectric and elastic effects contributing to the final response to the electric fields. 
The theoretical predictions are then used to analyse optically detected magnetic resonance (ODMR) spectra recorded on hBN crystals containing different densities of $\text{V}_\text{B}^{-}$ centres. We prove that the orthorhombic zero-field splitting parameter results from local electric fields produced by surrounding charge defects. 
By providing calculations of the spin-strain and spin-electric field couplings, this work paves the way towards applications of $\text{V}_\text{B}^{-}$ centres for quantitative electric field imaging and quantum sensing under pressure.
\end{abstract}

\maketitle
\newpage
\section{Introduction}
Electric and strain field effects are highly important ingredients in the tight control of solid-state spin defects for quantum technology applications. The accurate knowledge of the couplings to these perturbations can be harnessed either actively, e.g.\ for tuning the magneto-optical properties of the defect, or passively, e.g.\ for sensing its close environment. Particular attention is currently given to spin defects embedded in two-dimensional (2D) van der Waals materials~\cite{Toth_2019, Chakraborty_2019, Ye_2019,vaidya2023quantum}. The main advantage of a 2D host is the proximity of the quantum defect to the surface, which improves both its optical efficiency~\cite{Li_2019} and sensing capabilities~\cite{vaidya2023quantum}. One of these 2D materials is hexagonal boron nitride (hBN), which can be exfoliated down to the mono-layer limit without compromising its chemical stability~\cite{Song_2010, Park_2014}. Owing to its large optical band gap ($\sim 6.0~\mathrm{eV}$~\cite{Cassabois_2016}), hBN hosts optically-active point defects in a wide range of wavelengths. 
Once isolated at the individual scale, defects in hBN were first studied as single photon emitters, providing narrow and tunable emission lines, high brightness, and perfect photostability~\cite{Tran_2016_bulk, Martinez_2016, Vogl_2018, Li_2022,Jungwirth_2016, Grosso_2017, Noh_2018, Mendelson_2020,Sajid_2020_rev}. In addition, it was recently shown that the electron spin resonance frequencies of some defects can be inferred through optically detected magnetic resonance (ODMR) methods, providing a central resource for quantum sensing applications~\cite{Chejanovsky_2021, Stern_2022,Sajid_2018, Weston_2018, Sajid_2020, Auburger_2021,Gottscholl2020}. 

To date, the negatively-charged boron-vacancy centre ($\text{V}_\text{B}^{-}$) is the only point defect in hBN possessing an ODMR response whose microscopic structure has been unambiguously identified~\cite{Gottscholl2020, Abdi_2018, Ivady_2020, haykal2022}. This defect, which can be readily created by irradiation, ion-implantation and laser writing techniques~\cite{Li_2021, Murzakhanov_2021, Kianinia_2020, Guo_2022, Gao2021}, has been recently employed for sensing magnetic fields~\cite{Gottscholl2021, Gao_2021, Huang_2022, Kumar_2022, Healey_2023}, strain~\cite{Yang_2022,GaoStrain2022}, and temperature~\cite{Gottscholl2021, ACSPhot_Guo2021,Healey_2023}. In this context, the purpose of the present work is to provide an accurate description of the coupling of $\text{V}_\text{B}^{-}$ centres in hBN to strain and electric fields. 

Under optical illumination, the $\text{V}_\text{B}^{-}$ centre exhibits a very broad emission in the near-infrared~\cite{Gottscholl2020}. Although an experimental work has reported a zero-phonon-line (ZPL) at $773~\mathrm{nm}$~\cite{Qian2022}, a recent theoretical study has claimed that the full emission is phonon-assisted since optical transitions between electronic states are forbidden~\cite{Libbi_2022}. As a consequence, the optical emission is intrinsically dim and only dense ensembles of $\text{V}_\text{B}^{-}$ centres could be detected in experiments so far. Theory further identified that the $\text{V}_\text{B}^{-}$ centre has a spin triplet ground state ($S=1$) with $\text{D}_\text{3h}$ symmetry~\cite{Ivady_2020}. In the absence of external perturbations, the ground state spin Hamiltonian only involves spin-spin interaction and reads as
\begin{equation}\label{zfs}
    \hat{H}_\text{SS}=D\left(\hat{S}_{z}^2-\frac{S\left(S+1\right)}{3}\right) \ .
\end{equation}
The axial zero-field-splitting (ZFS) parameter $D$ separates the $m_S=0$ and $m_S=\pm1$ spin sublevels, the notation $m_S$ referring to the spin projection along the $c$-axis ($z$) of the hBN crystal. Interestingly, ODMR spectra recorded on ensembles of $\text{V}_\text{B}^{-}$ centres usually indicate two magnetic resonances at frequencies $D\pm E$, with $D\sim 3.47$~GHz and $E\sim 50$~MHz~\cite{Gottscholl2020}. These measurements thus reveal an additional orthorhombic splitting ($E$) of the $m_S=\pm1$ levels, which is in contradiction with the $\text{D}_\text{3h}$ symmetry of the $\text{V}_\text{B}^{-}$ centre. While this splitting was originally attributed to strain effects in the hBN crystal leading to a reduced symmetry of the defect~\cite{Gottscholl2020,Guo_2022}, a recent study suggested that it  results instead from the interaction with a local electric field~\cite{Gong_2022}. However, no microscopic picture has been applied to interpret these results. More generally, there is an urgent need to understand the coupling of strain and electric fields to $\text{V}_\text{B}^{-}$ defect spins, which is an exemplary quantum sensor in a van der Waals material. 

In this paper, we employ first principles simulations to determine the effect of external electric and strain fields on the ground state electronic spin structure of $\text{V}_\text{B}^{-}$ centres in hBN. We show that in most of the experimental conditions the defect experiences fluctuating electric fields that leads to an orthorhombic splitting in ODMR spectra recorded at zero external magnetic field. We identify the microscopic origin of this effect and determine the spin-strain and spin-electric field coupling parameters. We then compare these results to experimental ODMR spectra recorded from  neutron-irradiated hBN crystals containing different densities of $\text{V}_\text{B}^{-}$ centres. Our study proves that the orthorhombic ZFS is caused by fluctuating electric fields and suggest that the coupling strengths of the $\text{V}_\text{B}^{-}$ centres to strain and electric fields are comparable to those of the nitrogen-vacancy (NV$^-$) centre in diamond~\cite{Barson_2017, VanOort_1990}. 

\section{Results}
\subsection{Defect structure and spin-spin interaction}

\begin{figure}[t]
    \centering
    \includegraphics[width = 8.7cm]{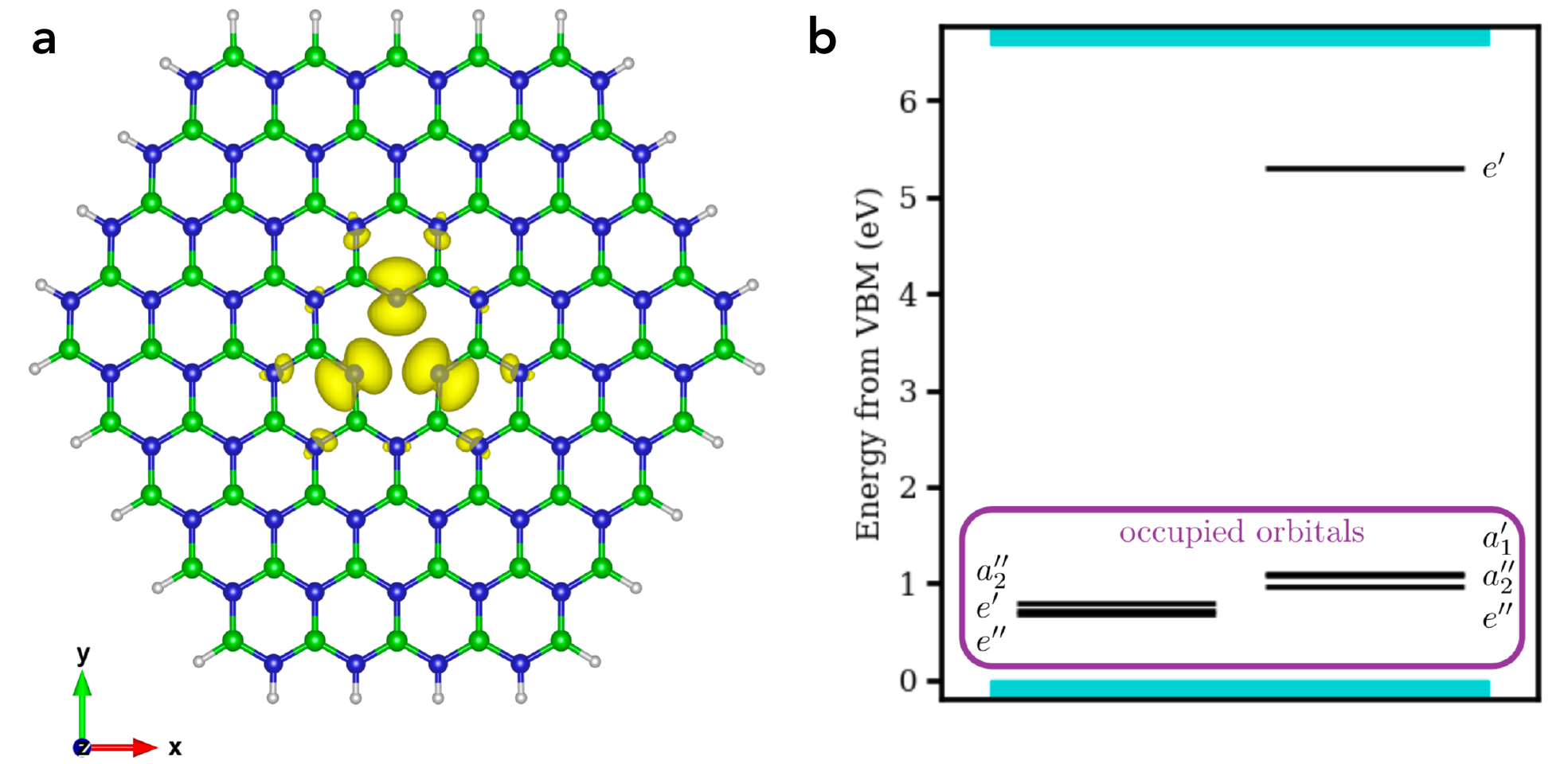}
    \caption{Geometry and electronic structure of the $\text{V}_\text{B}^{-}$ centre in hBN. {\bf a} Visualisation of the $\text{V}_\text{B}^{-}$ centre and its ground state electronic spin density (yellow) in the mono-layer flake model used in the calculations. Green, blue, and white balls represent boron, nitrogen and hydrogen atoms, respectively. {\bf b} Kohn-Sham level structure of the defect calculated with PBE0 functional. The orbitals are labelled according to the irreducible representations in $\text{D}_\text{3h}$ point group symmetry. The subspace of occupied orbitals are indicated in the figure. As these lie close to each other, their labels are ordered in ascending energy from bottom to top. The spin density of the triplet ground state originates from the unoccupied double-degenerate $e^{\prime}$ orbital.}
    \label{fig:geom}
\end{figure}

The $\text{V}_\text{B}^{-}$ centre in hBN is a single vacancy at the boron site in the negatively charged state possessing $\text{D}_{3\text{h}}$ point symmetry (Fig.~\ref{fig:geom}{\bf a}). In the following, the spin properties of the $\text{V}_\text{B}^{-}$ centre are calculated using a large hydrogen-terminated mono-layer flake model (see Methods). The coordinate frame is defined as $\{x,y\}$ span the plane of the hBN layer, with $x$ and $y$ along the zigzag and armchair directions of the hexagonal lattice, respectively. The Kohn-Sham levels in the spin-polarised calculations are plotted in Fig.~\ref{fig:geom}{\bf b}, showing nearby double degenerate $e^\prime$ and $e^{\prime\prime}$ levels and non-degenerate $a_1^\prime$ and $a_2^{\prime\prime}$ levels leaving two holes in the $e^\prime$ orbital of the spin minority spin channel. This constitutes the $^{3}A^{\prime}_{2}$ spin triplet ground state of the $\text{V}_\text{B}^{-}$ centre, in agreement with previous calculations~\cite{Abdi_2018, Ivady_2020, Sajid_2020_edge}. Importantly, the spin density is dominantly localised in-plane on the three dangling bonds of the neighbouring nitrogen atoms (Fig.~\ref{fig:geom}{\bf a}), as recently confirmed experimentally by pulsed electron-nuclear double resonance techniques~\cite{Gracheva2023}.

Starting with an unperturbed $\text{D}_{3\text{h}}$ symmetry, the mono-layer flake model leads to an axial ZFS parameter $D=3263~\mathrm{MHz}$ [see Eq.~\eqref{zfs}], a value in fair agreement with the experimental results $D\approx3470~\mathrm{MHz}$. In the Supplementary Note 1, we show that the mono-layer flake model well represents the bulk environment around the defect because the spin density is fully localised into the hBN plane. Although there is about 6\% discrepancy in the absolute value of the $D$ parameters obtained from our theoretical approach and the experimental data, the accuracy in the variation of the zero-field-splitting parameters upon external perturbations can be reliably obtained thanks to the cancellation of numerical errors in partial derivatives.

In the next sections, we compute the coupling coefficients of the $\text{V}_\text{B}^{-}$ centre to strain and electric fields. We assume that the strengths of theses fields are much smaller than the Coulomb interaction between the electrons, so that strain and electric fields can be considered as perturbations. We analyse how these interactions can lead to an orthorhombic ZFS parameter, that splits the $m_S=\pm1$ levels upon axial symmetry breaking. 

\subsection{Strain field interaction}

We start the analysis of external perturbations with the strain field case. We construct the interaction Hamiltonian analogously to previous calculations performed on the $\text{NV}^{-}$ centre in diamond featuring $\text{C}_\text{3v}$ symmetry~\cite{Udvarhelyi_2018}. In the linear approximation, the bi-linear forms of the second rank tensor of the external strain are connected to the bi-linear forms of the $\hat{S}\circ\hat{S}$ second rank tensor by scalar coupling coefficients $g$. We write the interaction in a symmetry-adapted form to retain only the linearly independent couplings, i.e. we construct the irreducible representation basis in $\text{D}_\text{3h}$ symmetry from the components of the strain and spin tensors. These transform as quadratic spatial coordinates, e.g. $zz$ and $xx+yy$ both belong to the totally symmetric $A_{1}^{\prime}$ irreducible representation. The interaction Hamiltonian is totally symmetric, thus only the products of the same irreducible representations of the strain and spin are valid combinations. 

According to the theory of invariants, the Hamiltonian describing the interaction of the $\text{V}_\text{B}^{-}$ centre with strain can be formulated as
\begin{align}\label{strain}
&\nonumber\hat{H}_\text{strain}=
\left(\varepsilon_{xx}+\varepsilon_{yy}\right)g_{1} \hat{S}_{z}^2 
\\\nonumber&+ \left(\varepsilon_{xx}-\varepsilon_{yy}\right) \left[g_{2}\left(\hat{S}_{x}^{2}-\hat{S}_{y}^{2}\right) +g^{\prime}_{2}\left(\hat{S}_{x}\hat{S}_{y}+\hat{S}_{y}\hat{S}_{x}\right)\right]
\\&+ \left(\varepsilon_{xy}+\varepsilon_{yx}\right) \left[g_{3}\left(\hat{S}_{x}^{2}-\hat{S}_{y}^{2}\right) +g^{\prime}_{3}\left(\hat{S}_{x}\hat{S}_{y}+\hat{S}_{y}\hat{S}_{x}\right)\right] 
\end{align}
where $g_i$ are the coupling coefficients between $\varepsilon$ bi-linear forms of the strain matrix elements and $\hat{S}$ bi-linear forms of the spin projection operators reflecting a symmetry-adapted form of $A^{\prime}_{1}\times A^{\prime}_{1} + E^{\prime} \times E^{\prime}$. Note that the out-of-plane strain components $\{\varepsilon_{zx},\varepsilon_{zy},\varepsilon_{zz}\}$ are not included in the Hamiltonian because we consider a single-layer flake model. However, these couplings are expected to be minor for a bulk model as well since out-of-plane ($z$) distortions have a small impact on the in-plane hBN structure that contains the spin density of the $\text{V}_\text{B}^{-}$ centre (see Supplementary Note~2). 

\begin{figure}[t]
    \centering
    \includegraphics[width = 8.7cm]{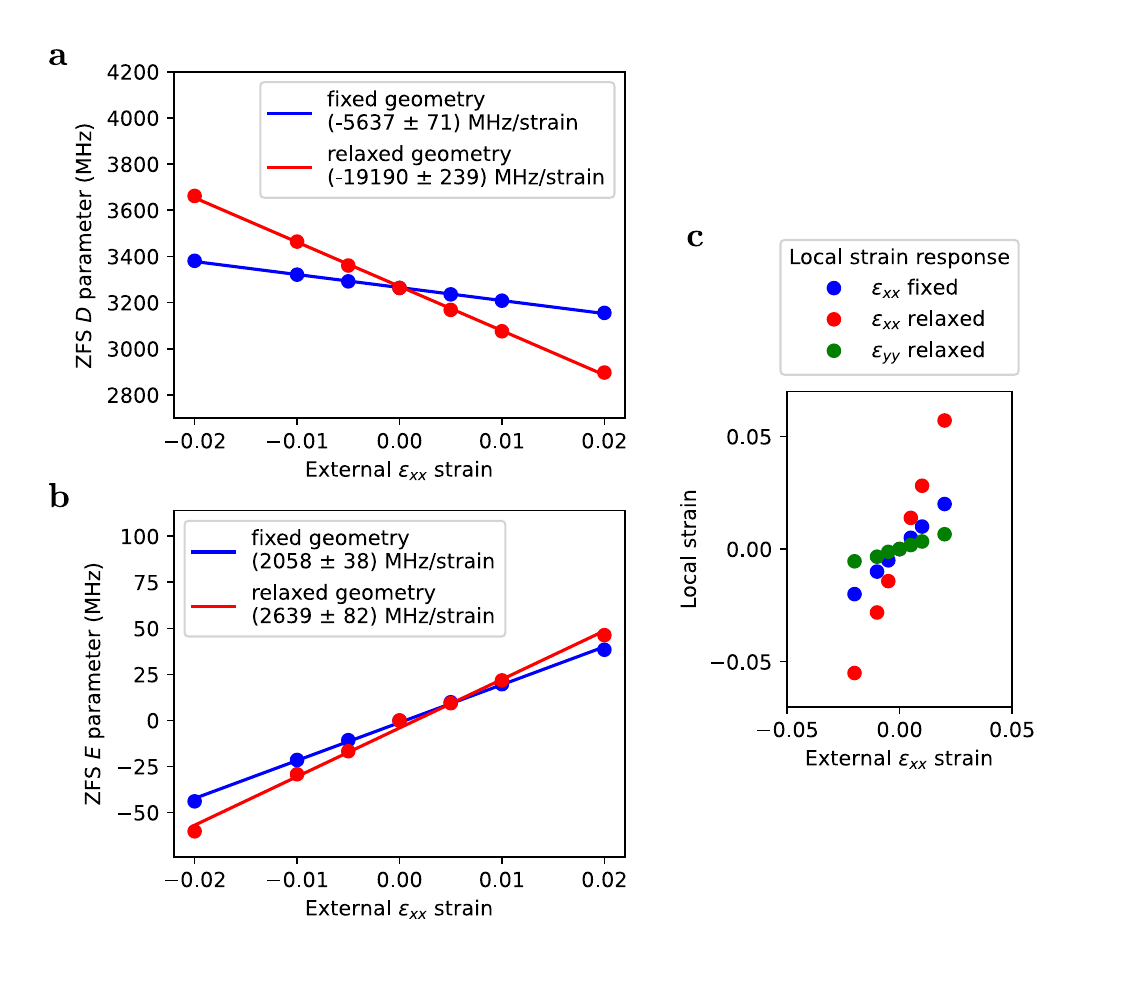}
    \caption{Response to an external strain component $\varepsilon_{xx}$. {\bf a}~Axial ZFS parameter $D$ as a function of the external strain. {\bf b}~Orthorhombic ZFS parameter $E$ as a function of the external strain. The cases of fixed and relaxed atomic positions are compared (blue and red points), as explained in the main text. {\bf c} Local strain response for the externally applied strain. Blue dots show the one-to-one correspondence when the atomic positions are not allowed to relax. Red and green points show the enhanced local $\varepsilon_{xx}$ strain and the activated local $\varepsilon_{yy}$ strain around the defect, respectively, when its close vicinity is allowed to relax under the strain applied to the crystal.}
    \label{fig:strain_xx}
\end{figure}

To calculate the coupling coefficients $g_i$, we specify uniform strain of a single element in the strain tensor in each calculation and apply it to our model (see Methods). We do so for each symmetrically non-equivalent components in Eq. \eqref{strain}, i.e. for $\varepsilon_{xx}$ and $\varepsilon_{xy}$. We then calculate the resulting axial and orthorhombic ZFS parameters (see red data in Fig.~\ref{fig:strain_xx}{\bf a},{\bf b}) and obtain the coupling coefficients $g_i$ as the partial derivatives
\begin{align}
g_{1}&=\frac{\partial D}{\partial \varepsilon_{xx}}= (-19.2\pm0.2)~\frac{\mathrm{GHz}}{\mathrm{strain}} \ \text{,}\\
g_{2}&=\frac{\partial (D_{xx}-D_{yy})/2}{\partial \varepsilon_{xx}}= (2.6\pm0.8) ~\frac{\mathrm{GHz}}{\mathrm{strain}}\ \text{,}\\
g^{\prime}_{2}&=\frac{\partial (D_{xy}+D_{yx})/2}{\partial \varepsilon_{xx}}= 0 \ \text{,}\\
g_{3}&=\frac{\partial (D_{xx}-D_{yy})/2}{\partial \varepsilon_{xy}}\approx 0 \ \text{,}\\
g^{\prime}_{3}&=\frac{\partial (D_{xy}+D_{yx})/2}{\partial \varepsilon_{xy}}= (5.8\pm0.1) ~\frac{\mathrm{GHz}}{\mathrm{strain}} \ \text{,}
\end{align}
where $D_{ij}$ are components of the ZFS $\bf D$-matrix defined by alternate form of the interaction $\hat{H}=\hat{\bf S}\bf D\hat{\bf S}$. The negative sign of $g_1$ implies a larger (smaller) axial ZFS parameter $D=3D_{zz}/2$ for compressive (tensile) external strain, in line with the decrease (increase) of the distance between the localised spin density lobes. 

We now analyse how the microscopic structure of the defect affects the response to the strain. To this end, we carried out  calculations for which we strained the lattice of the defective model without allowing any local ionic relaxation around the defect and then calculated the change in the ZFS parameters (blue points in Fig.~\ref{fig:strain_xx}{\bf a},{\bf b}). After allowing local relaxation under the externally applied strain (red points in Fig.~\ref{fig:strain_xx}{\bf a},{\bf b}), we can define the local geometry changes in the close vicinity of the defect. The local strain is defined from the deformation of the triangle spanned by the three neighbouring nitrogen atom positions. We plot this local strain as a function of the external $\varepsilon_{xx}$ strain in Fig.~\ref{fig:strain_xx}{\bf c}. We identify a large increase in the same component ($\varepsilon_{xx}$) of the local strain after relaxation. Moreover, an additional $\varepsilon_{yy}$ component is also activated. We attribute this effect to the smaller local stiffness at the vacancy site originating from the broken bonds. During the relaxation of the defect under external strain, it is energetically favourable for the vacancy site to accommodate an increased strain compared to the externally applied one, thus lowering the strain enthalpy of the host material in its vicinity. Generally, the same strain enhancement is expected in defects with dangling or weaker bonds than the ones in its host material, making this type of qubits more sensitive strain sensors. The increased local strain after relaxation reflects in the increase of the ZFS strain coupling strength as well. Note that relaxation under external sheer strain ($\varepsilon_{xy}$) shows similar trend in the ZFS coupling strength. 

Finally, we convert the strain-spin coupling strengths to stress couplings, as stress is usually more accessible and controllable in experiments than strain. Based on the experimentally available elastic parameters of $C_{11}=811~\mathrm{GPa}$ and $C_{12}=168~\mathrm{GPa}$ in bulk hBN~\cite{Bosak_2006}, we calculate the $h_{i}$ stress coupling coefficients as
\begin{align}
    h_{1}&=\frac{g_1}{C_{11}+C_{12}}= (-19.6\pm0.2)~\frac{\mathrm{MHz}}{\mathrm{GPa}}\\
    h_{2}&=\frac{g_{2}}{C_{11}-C_{12}}= (4.04\pm1.2) ~\frac{\mathrm{MHz}}{\mathrm{GPa}} \ ,
\end{align}
where $C_{ij}$ elements are the second order elastic constants (stiffness tensor) in the Voigt notation. The relation between the strain ($\varepsilon$) and stress ($\sigma$) can be then formulated in linear elasticity as
\begin{equation}
    \left(\begin{matrix}\sigma_{xx}\\\sigma_{yy}\end{matrix}\right)=\left(\begin{matrix}C_{11}&C_{12}\\C_{12}&C_{11}\end{matrix}\right)\left(\begin{matrix}\varepsilon_{xx}\\\varepsilon_{yy}\end{matrix}\right)\text{.}
\end{equation}
With the above conversion, the Hamiltonian takes a similar form to Eq.~\eqref{strain} using the stress coupling coefficients
\begin{align}\label{stress}
\hat{H}_\text{stress}&=h_{1}\left(\sigma_{xx}+\sigma_{yy}\right)\hat{S}_{z}^2 \\
&+ h_{2} \left(\sigma_{xx}-\sigma_{yy}\right)\left(\hat{S}_{x}^{2}-\hat{S}_{y}^{2}\right)\text{.}
\end{align}
\indent Importantly, our calculations already indicate that $|h_{1}| > |h_{2}|$, such that spin-stress coupling mostly leads to variations of the axial ZFS parameter $D$ without inducing a significant orthorhombic splitting. 
%Our results of $h_1=-19.6~\mathrm{MHz/GPa}$ can be compared to a previous calculations reporting $-7.9~\mathrm{MHz/GPa}$ for hydrostatic pressure coupling to the $D$ parameter~\cite{Sajid_2020_edge}. The same calculation attributed the non-zero $E$ parameter to strain effects at the edge of the sample. 
\subsection{Electric field interaction}

The Hamiltonian of the electric field coupling is constructed analogously to the strain coupling as
\begin{equation}\label{electric}
\hat{H}_\text{elec}=d_{\perp}\left[\mathcal{E}_{y}\left(\hat{S}_{x}^{2}-\hat{S}_{y}^{2}\right)+\mathcal{E}_{x}\left(\hat{S}_{x}\hat{S}_{y}+\hat{S}_{y}\hat{S}_{x}\right)\right]
\end{equation}
where $d_{\perp}$ is the coupling coefficient between $\mathcal{E}$ linear forms of the electric field vector and $\hat{S}$ bi-linear forms of the spin projection operators reflecting a symmetry-adapted form of $E^{\prime} \times E^{\prime}$. We note that the parallel component of the electric field ($\mathcal{E}_z$) with respect to the hBN plane normal vector belongs to the $A^{\prime\prime}_{2}$ irreducible representation, and it cannot couple to either $A^{\prime}_{1}$ or $E^{\prime}$ representations of the spin operators. As a result, the interaction with an electric field only introduces an orthorhombic splitting without inducing variations of the axial ZFS parameter $D$.
\begin{figure}[t]
    \centering
    \includegraphics[width=6.3cm]{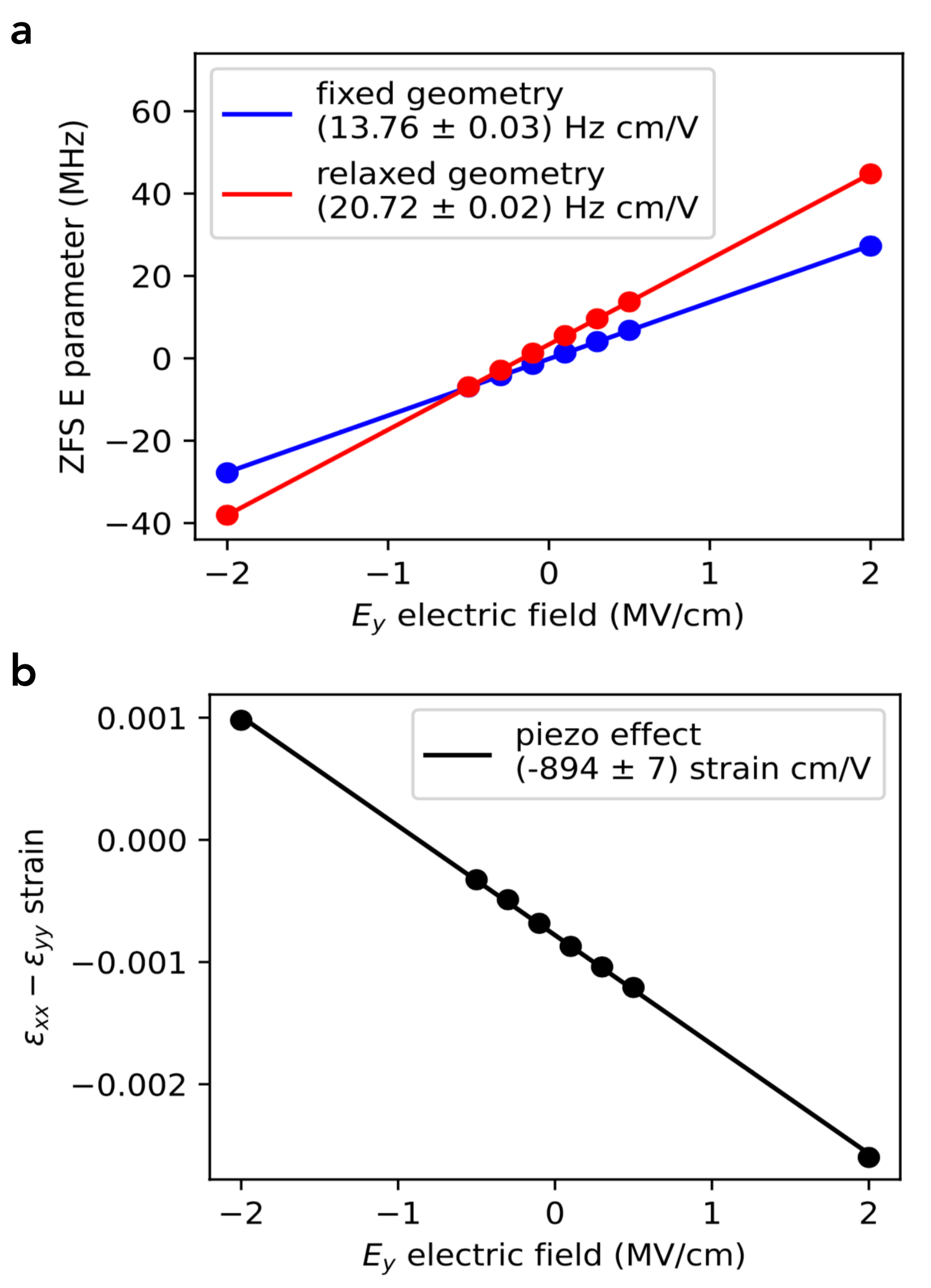}
    \caption{Response to an external electric field component $\mathcal{E}_{y}$. {\bf a} Orthorhombic ZFS parameter $E$ as a function of the external electric field. The cases of fixed and relaxed atomic positions are compared (blue and red points). {\bf b} Estimated local strain from the piezo effect induced by the external electric field.}
    \label{fig:electric}
\end{figure}

We apply homogeneous electric field to the defective hBN model.
The coupling coefficient can be calculated as the partial derivative of the ZFS parameter (see Fig.~\ref{fig:electric}{\bf a})
\begin{equation}
d_{\perp}=\frac{\partial E}{\partial \mathcal{E}_{y}}= (20.72\pm0.02)~\frac{\mathrm{Hz~cm}}{\mathrm{V}} \ .
\end{equation}
We note that a recent experimental work has estimated the coupling strength of the $E$ parameter to the electric field as $d_{\perp}=35~\mathrm{Hz \ cm/V}$~\cite{Gong_2022}, which is in the same order of magnitude as our first-principle results.

We now analyse the geometry relaxation under the applied external electric field similarly as for the case of the strain perturbation discussed above. We can identify an accompanying piezo effect resulting in $(\varepsilon_{xx}-\varepsilon_{yy})$ and $(\varepsilon_{xy}+\varepsilon_{yx})$ additional strains for the applied $\mathcal{E}_y$ and $\mathcal{E}_x$ electric fields, respectively. The former calculation is shown in Fig.~\ref{fig:electric}{\bf b}. This effect enhances the coupling strength to the electric field by a factor of $1.5$ (compare blue and red points in Fig.~\ref{fig:electric}{\bf a}).

The microscopic origin of the piezo enhancement effect can be explained by changes in electrostatic interactions in the vicinity of the defect under the applied external electric field. We obtain a significant change in the partial charge distribution of the strongly polarised bonds around the vacancy defect depending on the relative direction of the external electric field. This implies a redistribution of the electron density by the electric fields, which generates quantum mechanical forces on the ions. The origin of this effect can be detailed in the example of the vacancy-neighbour nitrogen atom in the $y$ direction and its boron neighbours. Without the external electric field applied, the B-N bonds are polarised with an effective dipole moment pointing from N to B. The applied field in the positive (resp. negative) direction increases (resp. decreases) the original charge separation and the effective dipole moment. This results in an additional effective attractive (resp. repulsive) force along the B-N bond and an outward (resp. inward) relaxation of the nitrogen atom relative to the vacancy site. We identify this additional relaxation as the $\varepsilon_{yy}$ piezo response. Thus, the relaxation of the nitrogen atom is in the same direction as the electric field. This finally enhances the symmetry-breaking effect of the electric field and consequently its coupling to the $E$ parameter.

\subsection{ODMR simulation}

\begin{figure}[t]
    \centering
    \includegraphics[width=8.7cm]{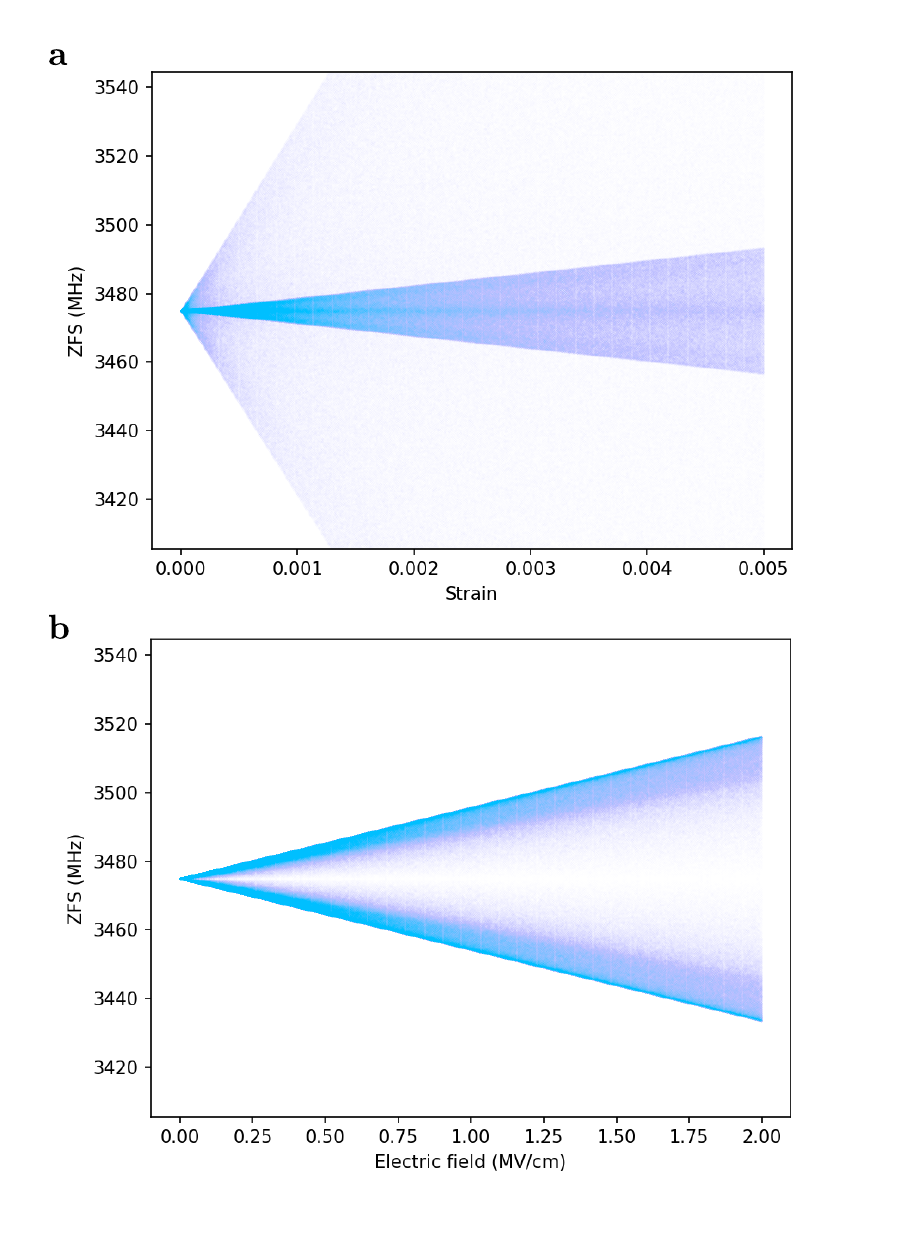}
    \caption{Simulated ZFS of the $\text{V}_\text{B}^{-}$ centre in random perturbation environments. {\bf a} ZFS distribution as a function of normal strain magnitude. {\bf b} ZFS distribution as a function of electric field magnitude.}
    \label{fig:sim}
\end{figure}

In the experimental ODMR signal of the $\text{V}_\text{B}^{-}$ centre, a small orthorhombic $E$ splitting was reported despite its contradiction with the defect symmetry. Recently, the effect was attributed to the local electric field originating from nearby charged defects without providing arguments from microscopic picture of the defect~\cite{Gong_2022}. Firstly, we simulate the ZFS parameters of a single $\text{V}_\text{B}^{-}$ defects affected by the presence of uniformly distributed random local strain or electric fields of given magnitudes. We sample 1000 random configurations for each given magnitude and visualise the corresponding $m_S=\pm1$ eigenvalues of the ZFS Hamiltonian as a scatter plot as a function of the perturbation magnitude. From the simulations shown in Fig.~\ref{fig:sim}, we conclude that the $E$ splitting can be exclusively attributed to the local electric field, while the effect of strain leads to a broadening of the signal. The sharp splitting originates from the large difference in the magnitudes of the couplings of the $D$ and $E$ parameters to the external fields. We can describe this with the ratio $g_D/g_E$ of the coupling strengths. In the case of the electric field, the $D$ parameter coupling is symmetry forbidden, $g_D/g_E\approx 0$, leading to a sharp splitting of the magnetic resonances. In the case of the strain field, we calculate $g_D/g_E=7.38$, which leads to a dominating shift in the ODMR signal with broadening of the integrated spectrum. We note that a similar effect was already observed and simulated for the $\text{NV}^{-}$ centre in diamond~\cite{Mittiga_2018}.

\begin{figure}[t!]
    \centering
    \includegraphics[width=8.7cm]{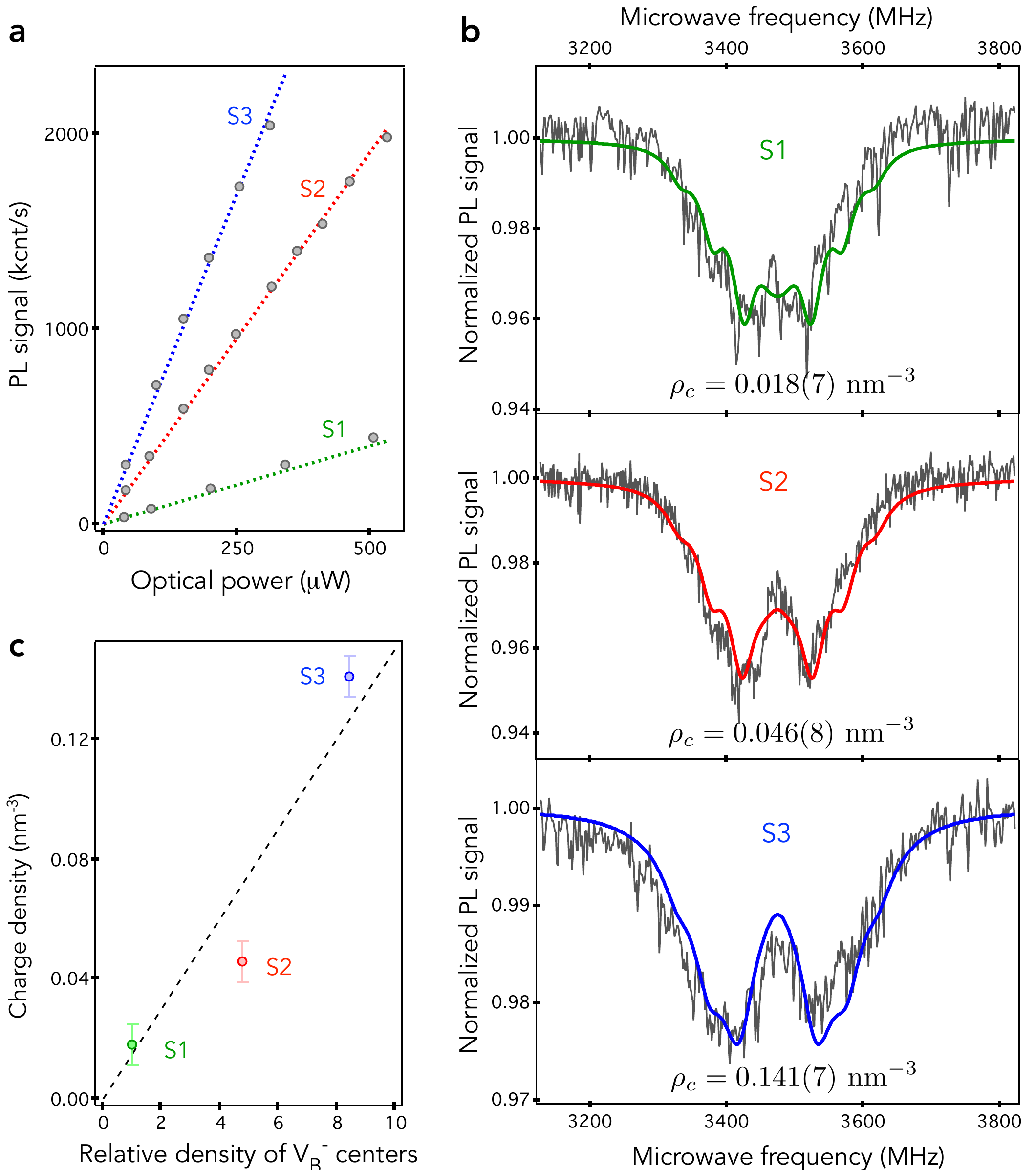}
    \caption{PL properties and ODMR spectra of hBN crystals with different concentration of $\text{V}_\text{B}^{-}$ centres. {\bf a} PL signal as a function of the optical excitation power for the three hBN crystals (S1, S2 and S3). The dashed lines are linear fit. {\bf b} ODMR spectra recorded for S1, S2 and S3 crystals from top to bottom. The solid lines are the results of the fitting procedure based on the microscopic charge model described in the main text, from which we extract the charge density $\rho_c$. {\bf c} Charge density $\rho_c$ as a function of the relative density of V$_{\rm B}^-$ centres in the hBN crystal. The dashed line is a linear fit.}
    \label{fig:ODMR}
\end{figure}

We now use our theoretical predictions to analyse experimental ODMR spectra recorded at zero external magnetic field. To this end, we rely on three hBN crystals (S1, S2 and S3) with different densities of V$_{\rm B}^-$ centres created by neutron irradiation. Their optical properties are analysed with a confocal microscope operating at room temperature under green laser illumination (see Methods). All crystals exhibit the characteristic emission spectrum of V$_{\rm B}^-$ centres with a broad spectral line in the near infrared. The relative density of V$_{\rm B}^-$ centres is estimated by recording the evolution of the PL signal with the optical excitation power [Fig.~\ref{fig:ODMR}{\bf a}]. In the considered power range, the PL signal increases linearly with a slope proportional to the density $\rho$ of V$_{\rm B}^-$ centres. By analysing these data, we obtain $\rho_{\rm S2}\sim 4.8\times\rho_{\rm S1}$ and $\rho_{\rm S3}\sim 8.5\times \rho_{\rm S1}$. 

The ODMR spectra recorded at zero external magnetic field are shown in Fig.~\ref{fig:ODMR}{\bf b}. For the three crystals, we detect the two characteristic magnetic resonances of the V$_{\rm B}^-$ centre with frequencies at $D\pm E$. Importantly, the $D$ parameter is identical for all spectra while the $E$-splitting increases with the density of V$_{\rm B}^-$ centre. To understand these results, we rely on a microscopic charge model originally introduced for $\text{NV}^{-}$ defects in diamond~\cite{Mittiga_2018}. We consider that negatively-charged V$_{\rm B}^-$ centres are associated to positively-charged defects in order to ensure charge neutrality of the hBN crystal. These charges produce a local electric field that depends on the charge density $\rho_c$. The Hamiltonian describing the interaction of a central V$_{\rm B}^-$ centre with this electric field is constructed from Eqs.~\eqref{zfs} and~\eqref{electric} using the calculated coupling coefficient $d_{\perp}$. Moreover, we include the hyperfine splittings from the first neighbour $^{14}\text{N}$ nuclear spins with  a coupling strength of $47~\mathrm{MHz}$~\cite{Ivady_2020, Liu2022}. We keep this value fixed in the simulation according to the obtained negligible effect of the external electric field on the hyperfine parameters. We note that the hyperfine constants are, however, sensitive to the strain (see Supplementary Note 3). For the microscopic model of the charge environment, a specific number of elementary point charges are placed at the atomic sites of hBN in a simulation radius of $10~\mathrm{nm}$, corresponding to an average charge density $\rho_c$ in the simulation sphere. Their position vectors relative to the $\text{V}_\text{B}^{-}$ defect at the origin of the sphere is sampled randomly from the standard normal distribution. The resulting total electric field at the origin is supplied to the Hamiltonian as an effective external electric field. The obtained spectrum is collected for $10^4$ different random charge configurations. We then optimise the charge density and the ODMR contrast to fit to the experimental spectra.

As shown in Fig.~\ref{fig:ODMR}{\bf b}, all ODMR spectra are well fitted by this microscopic charge model, leading to charge densities $\rho_{c,{\rm S1}}=0.018(3)$~nm$^{-3}$, $\rho_{c,{\rm S2}}=0.046(4)$~nm$^{-3}$ and $\rho_{c,{\rm S3}}=0.141(8)$~nm$^{-3}$. These results indicate that the density of charges increases linearly with the density of V$_{\rm B}^-$ centres in the hBN crystal (Fig.~\ref{fig:ODMR}{\bf c}), as expected from the microscopic charge model.

\section{Discussion}
We calculated the coupling strengths of the ZFS parameters to the symmetry adapted components of strain and electric fields for the $\text{V}_\text{B}^{-}$ defect in 2D hBN from first principles. In order to place these values in a broader context, we briefly compare them to those of the popular $\text{NV}^{-}$ defect in diamond. The calculated and experimental values of the stress coupling on the axial ZFS parameter of the $\text{NV}^{-}$ centre are $-5.17~\mathrm{MHz/GPa}$~\cite{Udvarhelyi_2018} and $-4.4~\mathrm{MHz/GPa}$~\cite{Barson_2017}, respectively. These values are weaker than that obtained for the $\text{V}_\text{B}^{-}$ centre $h_1=-19.6~\mathrm{MHz/GPa}$. On the other hand, the experimental perpendicular electric field coupling $d_{\perp}$ was reported to be $17~\mathrm{Hz\,cm/V}$ for $\text{NV}^{-}$ centres~\cite{VanOort_1990}, which is close to the result of our calculation of about $21~\mathrm{Hz\,cm/V}$ for the $\text{V}_\text{B}^{-}$ centre in hBN. 

For the strain perturbation, the coupling to the axial ZFS parameter $D$ dominates, while it is forbidden by symmetry for electric field perturbations. Consequently, we identify the electric field perturbation as the source of the symmetry-breaking orthorhombic $E$ splitting in the ODMR experiments. Our numerical simulations based on the calculated couplings can accurately model the experimental ODMR signal in the presence of electric field perturbations originating from point charges in the lattice sites of the bulk hBN host. Furthermore, we identify a correlation in the defect density created by neutron irradiation with the simulated density of point charges and ultimately with the effective $E$ splitting in the ODMR signal.

Moreover, our DFT calculations reveal a local enhancement effect in the coupling strengths for both strain and electric fields. The former is specific to the vacancy nature of the defect. The latter is attributed to the local piezo effect, i.e. an accompanying strain response under the external electric field effect. In other words, the local piezo effect strengthens the response of the ZFS to the external electric field which is a significant effect of about 50\%. We note that this effect goes to the opposite of one 3D qubit system, the divacancy qubit in 4H silicon carbide~\cite{Falk_2014}. In that case, the local piezo effect rather generates such quantum mechanical forces on the ions that compensate the effect of the external electric field and reduce the spin-electric field coupling by about 10\%. 

In summary, the spin-strain and spin-electric field couplings of V$_{\rm B}^-$ centre in hBN were determined. We proved that fluctuating electric fields around the V$_{\rm B}^-$ spins are at the origin of the orthorhombic splitting commonly observed in ODMR spectra of ensembles V$_{\rm B}^-$ centres at zero external magnetic field. We showed that the piezo effect is strong in the spin-electric field coupling parameters, which may be generalised for other vacancy-type defects in hBN. This work might guide future applications of V$_{\rm B}^-$ centres in hBN, such as quantum sensing under high pressure and electrometry.

\section{Methods}
\subsection{Density functional theory calculations}
The nano-flake model used in the calculation is a hydrogen terminated hBN mono-layer of up-to the fourth hexagon neighbours around the boron vacancy defect, consisting of 147 atoms in total. Its electronic structure was calculated by density functional theory (DFT) using the PBE0 functional~\cite{PBE0} and split valence polarisation Karlsruhe basis set (def2-SVP)~\cite{def2}, as implemented in the ORCA code~\cite{ORCA}. The spin-spin ZFS tensor is calculated using the restricted spin density obtained from the singly occupied unrestricted natural orbitals~\cite{Sinnecker_2006}.
During the relaxation of the defect structure under external strain applied, the outshell hBN hexagons were fixed in the atomic positions that are associated the strain environment acting to the inner part of the model.  

\subsection{Numerical simulations}
ZFS simulation of the defect under external strain and electric field perturbations is calculated in a home-built Python code. We iterate over the magnitude of the perturbation in $1000$ steps for the simulation range. In each iteration, we adaptively sample (scaled by the magnitude) the random configurations of effective perturbation vectors $\varepsilon=(\varepsilon_{xx}, \varepsilon_{yy})$ and $\mathcal{E}=(\mathcal{E}_x, \mathcal{E}_y)$ for the normal strain and electric fields, respectively. The component are sampled from uniform random distribution and the vector is normalised.
In the ODMR simulation, we sample a uniform random distribution of positions inside a sphere, representing parasitic charged defects around the target boron-vacancy defect. To this end, we sample the elements of the position vector from standard normal distribution and multiply with $d=r\sqrt[3]{x}$ random distance, where $x$ is sampled from a uniform random normalised distribution and $r$ is the simulation sphere radius. Then we define an underlying lattice of bulk hBN and project the positions to the nearest lattice sites. We optimise two parameters, the charge density and the ODMR contrast, to fit to the experimental spectrum. The optimisation is done by the least squares method using parallelized brute force method on a two dimensional parameter grid refined in three consecutive optimisation cycles. 

\subsection{hBN crystals}
The three hBN crystals used in this work were synthesised through the metal flux growth method described in Ref.~[\citenum{Liu2018}], and irradiated at the Ohio State University Research Reactor, which produces a thermal neutron flux of $10^{12}$~cm$^{-2}$s$^{-1}$. The crystal S1 is isotopically purified with $^{11}$B ($99.4\%$) and was irradiated with a neutron dose of $2.6 \times 10^{17}$~cm$^{-2}$. The two other crystals, S2 and S3, are purified with $^{10}$B ($99.2\%$) and irradiated with a dose of $2.6 \times 10^{16}$~cm$^{-2}$ and $2.6 \times 10^{17}$~cm$^{-2}$, respectively. Neutron irradiation creates V$_{\rm B}^-$ centres (i) through damages induced by neutron scattering across the crystal and (ii) via neutron absorption leading to nuclear transmutation doping. The latter process strongly depends on the isotopic content of the hBN crystal since the neutron capture cross-section of $^{10}$B is orders of magnitude larger than that of $^{11}$B. As a result, neutron irradiation creates more efficiently V$_{\rm B}^-$ centres in hBN crystals isotopically enriched with $^{10}$B.\\

\subsection{Experimental details}
The optical properties of hBN crystals are studied with a  confocal microscope operating at ambient conditions. A laser excitation at 532 nm is focused onto the sample with a high numerical aperture microscope objective (NA=0.95). The PL signal is collected by the same objective and directed to a silicon avalanche photodiode operating in the single-photon counting regime. ODMR spectra are recorded by monitoring the PL signal while sweeping the frequency of a microwave field applied through a copper microwire deposited on the hBN crystal surface.\\

\section*{Declarations}
\begin{itemize}
 \item {\bf Data Availability} The authors declare that the main data supporting the findings of this study are available within the paper and its Supplementary files. Part of source data is provided with this paper. The data that support the findings of this study are available from the corresponding author upon reasonable request.
 
 \item {\bf Code Availability} The codes that were used in this study are available upon request to the corresponding author.

\item {\bf Acknowledgement} This work was supported by the National Excellence Program for the project of Quantum-coherent materials (NKFIH Grant No. KKP129866), the Ministry of Culture and Innovation and the National Research, Development and Innovation Office within the Quantum Information National Laboratory of Hungary (Grant No. 2022-2.1.1-NL-2022-00004), the French Agence Nationale de la Recherche under the program ESR/EquipEx+ (grant number ANR-21-ESRE-0025), the Institute for Quantum Technologies in Occitanie through the project BONIQs and Qfoil. Support for hBN crystal growth is provided by the Office of Naval Research, awards numbers N00014-22-1-2582 and N00014-20-1-2474.  The neutron irradiation was supported by the U.S. Department of Energy, Office of Nuclear Energy under DOE Idaho Operations Office Contract DE-AC07- 051D14517 as part of a Nuclear Science User Facilities experiment. We acknowledge the support of The Ohio State University Nuclear Reactor Laboratory and the assistance of Susan M. White, Lei Raymond Cao, Andrew Kauffman, and Kevin Herminghuysen for the irradiation services provided. We acknowledge the high-performance computational resources provided by KIFÜ (Governmental Agency for IT Development) institute of Hungary. 

\item {\bf Funding} Open access funding provided by ELKH Wigner Research Centre for Physics. 

\item {\bf Author contribution} PU carried out the DFT calculations and analysed the results under the supervision of AG. TCP, AD, BG, GC and VJ performed the experiments. JL and JHE provided the neutron-irradiated hBN crystals. PU, VJ and AG wrote the paper. All authors contributed to the discussion and commented on the manuscript. AG conceived and led the entire scientific project.

\item {\bf Competing Interests} The authors declare that there are no competing interests.

\item {\bf Correspondence} Correspondence should be addressed to A.G.~(email: gali.adam@wigner.hu).
\end{itemize}

\bibliography{bib}

%merlin.mbs apsrev4-1.bst 2010-07-25 4.21a (PWD, AO, DPC) hacked
%Control: key (0)
%Control: author (72) initials jnrlst
%Control: editor formatted (1) identically to author
%Control: production of article title (-1) disabled
%Control: page (0) single
%Control: year (1) truncated
%Control: production of eprint (0) enabled
\begin{thebibliography}{57}%
\makeatletter
\providecommand \@ifxundefined [1]{%
 \@ifx{#1\undefined}
}%
\providecommand \@ifnum [1]{%
 \ifnum #1\expandafter \@firstoftwo
 \else \expandafter \@secondoftwo
 \fi
}%
\providecommand \@ifx [1]{%
 \ifx #1\expandafter \@firstoftwo
 \else \expandafter \@secondoftwo
 \fi
}%
\providecommand \natexlab [1]{#1}%
\providecommand \enquote  [1]{``#1''}%
\providecommand \bibnamefont  [1]{#1}%
\providecommand \bibfnamefont [1]{#1}%
\providecommand \citenamefont [1]{#1}%
\providecommand \href@noop [0]{\@secondoftwo}%
\providecommand \href [0]{\begingroup \@sanitize@url \@href}%
\providecommand \@href[1]{\@@startlink{#1}\@@href}%
\providecommand \@@href[1]{\endgroup#1\@@endlink}%
\providecommand \@sanitize@url [0]{\catcode `\\12\catcode `\$12\catcode
  `\&12\catcode `\#12\catcode `\^12\catcode `\_12\catcode `\%12\relax}%
\providecommand \@@startlink[1]{}%
\providecommand \@@endlink[0]{}%
\providecommand \url  [0]{\begingroup\@sanitize@url \@url }%
\providecommand \@url [1]{\endgroup\@href {#1}{\urlprefix }}%
\providecommand \urlprefix  [0]{URL }%
\providecommand \Eprint [0]{\href }%
\providecommand \doibase [0]{http://dx.doi.org/}%
\providecommand \selectlanguage [0]{\@gobble}%
\providecommand \bibinfo  [0]{\@secondoftwo}%
\providecommand \bibfield  [0]{\@secondoftwo}%
\providecommand \translation [1]{[#1]}%
\providecommand \BibitemOpen [0]{}%
\providecommand \bibitemStop [0]{}%
\providecommand \bibitemNoStop [0]{.\EOS\space}%
\providecommand \EOS [0]{\spacefactor3000\relax}%
\providecommand \BibitemShut  [1]{\csname bibitem#1\endcsname}%
\let\auto@bib@innerbib\@empty
%</preamble>
\bibitem [{\citenamefont {Toth}\ and\ \citenamefont
  {Aharonovich}(2019)}]{Toth_2019}%
  \BibitemOpen
  \bibfield  {author} {\bibinfo {author} {\bibfnamefont {M.}~\bibnamefont
  {Toth}}\ and\ \bibinfo {author} {\bibfnamefont {I.}~\bibnamefont
  {Aharonovich}},\ }\href
  {https://doi.org/10.1146/annurev-physchem-042018-052628} {\bibfield
  {journal} {\bibinfo  {journal} {Annual Review of Physical Chemistry}\
  }\textbf {\bibinfo {volume} {70}},\ \bibinfo {pages} {123} (\bibinfo {year}
  {2019})}\BibitemShut {NoStop}%
\bibitem [{\citenamefont {Chakraborty}\ \emph {et~al.}(2019)\citenamefont
  {Chakraborty}, \citenamefont {Vamivakas},\ and\ \citenamefont
  {Englund}}]{Chakraborty_2019}%
  \BibitemOpen
  \bibfield  {author} {\bibinfo {author} {\bibfnamefont {C.}~\bibnamefont
  {Chakraborty}}, \bibinfo {author} {\bibfnamefont {N.}~\bibnamefont
  {Vamivakas}}, \ and\ \bibinfo {author} {\bibfnamefont {D.}~\bibnamefont
  {Englund}},\ }\href@noop {} {\bibfield  {journal} {\bibinfo  {journal}
  {Nanophotonics}\ }\textbf {\bibinfo {volume} {8}},\ \bibinfo {pages} {2017}
  (\bibinfo {year} {2019})}\BibitemShut {NoStop}%
\bibitem [{\citenamefont {Ye}\ \emph {et~al.}(2019)\citenamefont {Ye},
  \citenamefont {Seo},\ and\ \citenamefont {Galli}}]{Ye_2019}%
  \BibitemOpen
  \bibfield  {author} {\bibinfo {author} {\bibfnamefont {M.}~\bibnamefont
  {Ye}}, \bibinfo {author} {\bibfnamefont {H.}~\bibnamefont {Seo}}, \ and\
  \bibinfo {author} {\bibfnamefont {G.}~\bibnamefont {Galli}},\ }\href@noop {}
  {\bibfield  {journal} {\bibinfo  {journal} {npj Computational Materials}\
  }\textbf {\bibinfo {volume} {5}},\ \bibinfo {pages} {44} (\bibinfo {year}
  {2019})}\BibitemShut {NoStop}%
\bibitem [{\citenamefont {Vaidya}\ \emph {et~al.}(2023)\citenamefont {Vaidya},
  \citenamefont {Gao}, \citenamefont {Dikshit}, \citenamefont {Aharonovich},\
  and\ \citenamefont {Li}}]{vaidya2023quantum}%
  \BibitemOpen
  \bibfield  {author} {\bibinfo {author} {\bibfnamefont {S.}~\bibnamefont
  {Vaidya}}, \bibinfo {author} {\bibfnamefont {X.}~\bibnamefont {Gao}},
  \bibinfo {author} {\bibfnamefont {S.}~\bibnamefont {Dikshit}}, \bibinfo
  {author} {\bibfnamefont {I.}~\bibnamefont {Aharonovich}}, \ and\ \bibinfo
  {author} {\bibfnamefont {T.}~\bibnamefont {Li}},\ }\href@noop {} {\enquote
  {\bibinfo {title} {Quantum sensing and imaging with spin defects in hexagonal
  boron nitride},}\ } (\bibinfo {year} {2023}),\ \Eprint
  {http://arxiv.org/abs/2302.11169} {arXiv:2302.11169} \BibitemShut {NoStop}%
\bibitem [{\citenamefont {Li}\ \emph {et~al.}(2019)\citenamefont {Li},
  \citenamefont {Scully}, \citenamefont {Shayan}, \citenamefont {Luo},\ and\
  \citenamefont {Strauf}}]{Li_2019}%
  \BibitemOpen
  \bibfield  {author} {\bibinfo {author} {\bibfnamefont {X.}~\bibnamefont
  {Li}}, \bibinfo {author} {\bibfnamefont {R.~A.}\ \bibnamefont {Scully}},
  \bibinfo {author} {\bibfnamefont {K.}~\bibnamefont {Shayan}}, \bibinfo
  {author} {\bibfnamefont {Y.}~\bibnamefont {Luo}}, \ and\ \bibinfo {author}
  {\bibfnamefont {S.}~\bibnamefont {Strauf}},\ }\href
  {https://doi.org/10.1021/acsnano.9b01996} {\bibfield  {journal} {\bibinfo
  {journal} {ACS Nano}\ }\textbf {\bibinfo {volume} {13}},\ \bibinfo {pages}
  {6992} (\bibinfo {year} {2019})}\BibitemShut {NoStop}%
\bibitem [{\citenamefont {Song}\ \emph {et~al.}(2010)\citenamefont {Song},
  \citenamefont {Ci}, \citenamefont {Lu}, \citenamefont {Sorokin},
  \citenamefont {Jin}, \citenamefont {Ni}, \citenamefont {Kvashnin},
  \citenamefont {Kvashnin}, \citenamefont {Lou}, \citenamefont {Yakobson},\
  and\ \citenamefont {Ajayan}}]{Song_2010}%
  \BibitemOpen
  \bibfield  {author} {\bibinfo {author} {\bibfnamefont {L.}~\bibnamefont
  {Song}}, \bibinfo {author} {\bibfnamefont {L.}~\bibnamefont {Ci}}, \bibinfo
  {author} {\bibfnamefont {H.}~\bibnamefont {Lu}}, \bibinfo {author}
  {\bibfnamefont {P.~B.}\ \bibnamefont {Sorokin}}, \bibinfo {author}
  {\bibfnamefont {C.}~\bibnamefont {Jin}}, \bibinfo {author} {\bibfnamefont
  {J.}~\bibnamefont {Ni}}, \bibinfo {author} {\bibfnamefont {A.~G.}\
  \bibnamefont {Kvashnin}}, \bibinfo {author} {\bibfnamefont {D.~G.}\
  \bibnamefont {Kvashnin}}, \bibinfo {author} {\bibfnamefont {J.}~\bibnamefont
  {Lou}}, \bibinfo {author} {\bibfnamefont {B.~I.}\ \bibnamefont {Yakobson}}, \
  and\ \bibinfo {author} {\bibfnamefont {P.~M.}\ \bibnamefont {Ajayan}},\
  }\href {https://doi.org/10.1021/nl1022139} {\bibfield  {journal} {\bibinfo
  {journal} {Nano Letters}\ }\textbf {\bibinfo {volume} {10}},\ \bibinfo
  {pages} {3209} (\bibinfo {year} {2010})}\BibitemShut {NoStop}%
\bibitem [{\citenamefont {Park}\ \emph {et~al.}(2014)\citenamefont {Park},
  \citenamefont {Park}, \citenamefont {Yun}, \citenamefont {Kim}, \citenamefont
  {Luong}, \citenamefont {Kim}, \citenamefont {Choi}, \citenamefont {Yang},
  \citenamefont {Kong}, \citenamefont {Kim},\ and\ \citenamefont
  {Lee}}]{Park_2014}%
  \BibitemOpen
  \bibfield  {author} {\bibinfo {author} {\bibfnamefont {J.-H.}\ \bibnamefont
  {Park}}, \bibinfo {author} {\bibfnamefont {J.~C.}\ \bibnamefont {Park}},
  \bibinfo {author} {\bibfnamefont {S.~J.}\ \bibnamefont {Yun}}, \bibinfo
  {author} {\bibfnamefont {H.}~\bibnamefont {Kim}}, \bibinfo {author}
  {\bibfnamefont {D.~H.}\ \bibnamefont {Luong}}, \bibinfo {author}
  {\bibfnamefont {S.~M.}\ \bibnamefont {Kim}}, \bibinfo {author} {\bibfnamefont
  {S.~H.}\ \bibnamefont {Choi}}, \bibinfo {author} {\bibfnamefont
  {W.}~\bibnamefont {Yang}}, \bibinfo {author} {\bibfnamefont {J.}~\bibnamefont
  {Kong}}, \bibinfo {author} {\bibfnamefont {K.~K.}\ \bibnamefont {Kim}}, \
  and\ \bibinfo {author} {\bibfnamefont {Y.~H.}\ \bibnamefont {Lee}},\ }\href
  {https://doi.org/10.1021/nn503140y} {\bibfield  {journal} {\bibinfo
  {journal} {ACS Nano}\ }\textbf {\bibinfo {volume} {8}},\ \bibinfo {pages}
  {8520} (\bibinfo {year} {2014})}\BibitemShut {NoStop}%
\bibitem [{\citenamefont {Cassabois}\ \emph {et~al.}(2016)\citenamefont
  {Cassabois}, \citenamefont {Valvin},\ and\ \citenamefont
  {Gil}}]{Cassabois_2016}%
  \BibitemOpen
  \bibfield  {author} {\bibinfo {author} {\bibfnamefont {G.}~\bibnamefont
  {Cassabois}}, \bibinfo {author} {\bibfnamefont {P.}~\bibnamefont {Valvin}}, \
  and\ \bibinfo {author} {\bibfnamefont {B.}~\bibnamefont {Gil}},\ }\href
  {\doibase 10.1038/nphoton.2015.277} {\bibfield  {journal} {\bibinfo
  {journal} {Nature Photonics}\ }\textbf {\bibinfo {volume} {10}},\ \bibinfo
  {pages} {262} (\bibinfo {year} {2016})}\BibitemShut {NoStop}%
\bibitem [{\citenamefont {Tran}\ \emph {et~al.}(2016)\citenamefont {Tran},
  \citenamefont {Zachreson}, \citenamefont {Berhane}, \citenamefont {Bray},
  \citenamefont {Sandstrom}, \citenamefont {Li}, \citenamefont {Taniguchi},
  \citenamefont {Watanabe}, \citenamefont {Aharonovich},\ and\ \citenamefont
  {Toth}}]{Tran_2016_bulk}%
  \BibitemOpen
  \bibfield  {author} {\bibinfo {author} {\bibfnamefont {T.~T.}\ \bibnamefont
  {Tran}}, \bibinfo {author} {\bibfnamefont {C.}~\bibnamefont {Zachreson}},
  \bibinfo {author} {\bibfnamefont {A.~M.}\ \bibnamefont {Berhane}}, \bibinfo
  {author} {\bibfnamefont {K.}~\bibnamefont {Bray}}, \bibinfo {author}
  {\bibfnamefont {R.~G.}\ \bibnamefont {Sandstrom}}, \bibinfo {author}
  {\bibfnamefont {L.~H.}\ \bibnamefont {Li}}, \bibinfo {author} {\bibfnamefont
  {T.}~\bibnamefont {Taniguchi}}, \bibinfo {author} {\bibfnamefont
  {K.}~\bibnamefont {Watanabe}}, \bibinfo {author} {\bibfnamefont
  {I.}~\bibnamefont {Aharonovich}}, \ and\ \bibinfo {author} {\bibfnamefont
  {M.}~\bibnamefont {Toth}},\ }\href {\doibase 10.1103/PhysRevApplied.5.034005}
  {\bibfield  {journal} {\bibinfo  {journal} {Phys. Rev. Appl.}\ }\textbf
  {\bibinfo {volume} {5}},\ \bibinfo {pages} {034005} (\bibinfo {year}
  {2016})}\BibitemShut {NoStop}%
\bibitem [{\citenamefont {Mart\'{\i}nez}\ \emph {et~al.}(2016)\citenamefont
  {Mart\'{\i}nez}, \citenamefont {Pelini}, \citenamefont {Waselowski},
  \citenamefont {Maze}, \citenamefont {Gil}, \citenamefont {Cassabois},\ and\
  \citenamefont {Jacques}}]{Martinez_2016}%
  \BibitemOpen
  \bibfield  {author} {\bibinfo {author} {\bibfnamefont {L.~J.}\ \bibnamefont
  {Mart\'{\i}nez}}, \bibinfo {author} {\bibfnamefont {T.}~\bibnamefont
  {Pelini}}, \bibinfo {author} {\bibfnamefont {V.}~\bibnamefont {Waselowski}},
  \bibinfo {author} {\bibfnamefont {J.~R.}\ \bibnamefont {Maze}}, \bibinfo
  {author} {\bibfnamefont {B.}~\bibnamefont {Gil}}, \bibinfo {author}
  {\bibfnamefont {G.}~\bibnamefont {Cassabois}}, \ and\ \bibinfo {author}
  {\bibfnamefont {V.}~\bibnamefont {Jacques}},\ }\href {\doibase
  10.1103/PhysRevB.94.121405} {\bibfield  {journal} {\bibinfo  {journal} {Phys.
  Rev. B}\ }\textbf {\bibinfo {volume} {94}},\ \bibinfo {pages} {121405}
  (\bibinfo {year} {2016})}\BibitemShut {NoStop}%
\bibitem [{\citenamefont {Vogl}\ \emph {et~al.}(2018)\citenamefont {Vogl},
  \citenamefont {Campbell}, \citenamefont {Buchler}, \citenamefont {Lu},\ and\
  \citenamefont {Lam}}]{Vogl_2018}%
  \BibitemOpen
  \bibfield  {author} {\bibinfo {author} {\bibfnamefont {T.}~\bibnamefont
  {Vogl}}, \bibinfo {author} {\bibfnamefont {G.}~\bibnamefont {Campbell}},
  \bibinfo {author} {\bibfnamefont {B.~C.}\ \bibnamefont {Buchler}}, \bibinfo
  {author} {\bibfnamefont {Y.}~\bibnamefont {Lu}}, \ and\ \bibinfo {author}
  {\bibfnamefont {P.~K.}\ \bibnamefont {Lam}},\ }\href
  {https://doi.org/10.1021/acsphotonics.8b00127} {\bibfield  {journal}
  {\bibinfo  {journal} {ACS Photonics}\ }\textbf {\bibinfo {volume} {5}},\
  \bibinfo {pages} {2305} (\bibinfo {year} {2018})}\BibitemShut {NoStop}%
\bibitem [{\citenamefont {Li}\ \emph {et~al.}(2022)\citenamefont {Li},
  \citenamefont {Pershin}, \citenamefont {Thiering}, \citenamefont
  {Udvarhelyi},\ and\ \citenamefont {Gali}}]{Li_2022}%
  \BibitemOpen
  \bibfield  {author} {\bibinfo {author} {\bibfnamefont {S.}~\bibnamefont
  {Li}}, \bibinfo {author} {\bibfnamefont {A.}~\bibnamefont {Pershin}},
  \bibinfo {author} {\bibfnamefont {G.}~\bibnamefont {Thiering}}, \bibinfo
  {author} {\bibfnamefont {P.}~\bibnamefont {Udvarhelyi}}, \ and\ \bibinfo
  {author} {\bibfnamefont {A.}~\bibnamefont {Gali}},\ }\href
  {https://doi.org/10.1021/acs.jpclett.2c00665} {\bibfield  {journal} {\bibinfo
   {journal} {The Journal of Physical Chemistry Letters}\ }\textbf {\bibinfo
  {volume} {13}},\ \bibinfo {pages} {3150} (\bibinfo {year}
  {2022})}\BibitemShut {NoStop}%
\bibitem [{\citenamefont {Jungwirth}\ \emph {et~al.}(2016)\citenamefont
  {Jungwirth}, \citenamefont {Calderon}, \citenamefont {Ji}, \citenamefont
  {Spencer}, \citenamefont {Flatté},\ and\ \citenamefont
  {Fuchs}}]{Jungwirth_2016}%
  \BibitemOpen
  \bibfield  {author} {\bibinfo {author} {\bibfnamefont {N.~R.}\ \bibnamefont
  {Jungwirth}}, \bibinfo {author} {\bibfnamefont {B.}~\bibnamefont {Calderon}},
  \bibinfo {author} {\bibfnamefont {Y.}~\bibnamefont {Ji}}, \bibinfo {author}
  {\bibfnamefont {M.~G.}\ \bibnamefont {Spencer}}, \bibinfo {author}
  {\bibfnamefont {M.~E.}\ \bibnamefont {Flatté}}, \ and\ \bibinfo {author}
  {\bibfnamefont {G.~D.}\ \bibnamefont {Fuchs}},\ }\href
  {https://doi.org/10.1021/acs.nanolett.6b01987} {\bibfield  {journal}
  {\bibinfo  {journal} {Nano Letters}\ }\textbf {\bibinfo {volume} {16}},\
  \bibinfo {pages} {6052} (\bibinfo {year} {2016})}\BibitemShut {NoStop}%
\bibitem [{\citenamefont {Grosso}\ \emph {et~al.}(2017)\citenamefont {Grosso},
  \citenamefont {Moon}, \citenamefont {Lienhard}, \citenamefont {Ali},
  \citenamefont {Efetov}, \citenamefont {Furchi}, \citenamefont
  {Jarillo-Herrero}, \citenamefont {Ford}, \citenamefont {Aharonovich},\ and\
  \citenamefont {Englund}}]{Grosso_2017}%
  \BibitemOpen
  \bibfield  {author} {\bibinfo {author} {\bibfnamefont {G.}~\bibnamefont
  {Grosso}}, \bibinfo {author} {\bibfnamefont {H.}~\bibnamefont {Moon}},
  \bibinfo {author} {\bibfnamefont {B.}~\bibnamefont {Lienhard}}, \bibinfo
  {author} {\bibfnamefont {S.}~\bibnamefont {Ali}}, \bibinfo {author}
  {\bibfnamefont {D.~K.}\ \bibnamefont {Efetov}}, \bibinfo {author}
  {\bibfnamefont {M.~M.}\ \bibnamefont {Furchi}}, \bibinfo {author}
  {\bibfnamefont {P.}~\bibnamefont {Jarillo-Herrero}}, \bibinfo {author}
  {\bibfnamefont {M.~J.}\ \bibnamefont {Ford}}, \bibinfo {author}
  {\bibfnamefont {I.}~\bibnamefont {Aharonovich}}, \ and\ \bibinfo {author}
  {\bibfnamefont {D.}~\bibnamefont {Englund}},\ }\href {\doibase
  10.1038/s41467-017-00810-2} {\bibfield  {journal} {\bibinfo  {journal}
  {Nature Communications}\ }\textbf {\bibinfo {volume} {8}},\ \bibinfo {pages}
  {705} (\bibinfo {year} {2017})}\BibitemShut {NoStop}%
\bibitem [{\citenamefont {Noh}\ \emph {et~al.}(2018)\citenamefont {Noh},
  \citenamefont {Choi}, \citenamefont {Kim}, \citenamefont {Im}, \citenamefont
  {Kim}, \citenamefont {Seo},\ and\ \citenamefont {Lee}}]{Noh_2018}%
  \BibitemOpen
  \bibfield  {author} {\bibinfo {author} {\bibfnamefont {G.}~\bibnamefont
  {Noh}}, \bibinfo {author} {\bibfnamefont {D.}~\bibnamefont {Choi}}, \bibinfo
  {author} {\bibfnamefont {J.-H.}\ \bibnamefont {Kim}}, \bibinfo {author}
  {\bibfnamefont {D.-G.}\ \bibnamefont {Im}}, \bibinfo {author} {\bibfnamefont
  {Y.-H.}\ \bibnamefont {Kim}}, \bibinfo {author} {\bibfnamefont
  {H.}~\bibnamefont {Seo}}, \ and\ \bibinfo {author} {\bibfnamefont
  {J.}~\bibnamefont {Lee}},\ }\href
  {https://doi.org/10.1021/acs.nanolett.8b01030} {\bibfield  {journal}
  {\bibinfo  {journal} {Nano Letters}\ }\textbf {\bibinfo {volume} {18}},\
  \bibinfo {pages} {4710} (\bibinfo {year} {2018})}\BibitemShut {NoStop}%
\bibitem [{\citenamefont {Mendelson}\ \emph {et~al.}(2020)\citenamefont
  {Mendelson}, \citenamefont {Doherty}, \citenamefont {Toth}, \citenamefont
  {Aharonovich},\ and\ \citenamefont {Tran}}]{Mendelson_2020}%
  \BibitemOpen
  \bibfield  {author} {\bibinfo {author} {\bibfnamefont {N.}~\bibnamefont
  {Mendelson}}, \bibinfo {author} {\bibfnamefont {M.}~\bibnamefont {Doherty}},
  \bibinfo {author} {\bibfnamefont {M.}~\bibnamefont {Toth}}, \bibinfo {author}
  {\bibfnamefont {I.}~\bibnamefont {Aharonovich}}, \ and\ \bibinfo {author}
  {\bibfnamefont {T.~T.}\ \bibnamefont {Tran}},\ }\href
  {https://onlinelibrary.wiley.com/doi/abs/10.1002/adma.201908316} {\bibfield
  {journal} {\bibinfo  {journal} {Advanced Materials}\ }\textbf {\bibinfo
  {volume} {32}},\ \bibinfo {pages} {1908316} (\bibinfo {year}
  {2020})}\BibitemShut {NoStop}%
\bibitem [{\citenamefont {Sajid}\ \emph
  {et~al.}(2020{\natexlab{a}})\citenamefont {Sajid}, \citenamefont {Ford},\
  and\ \citenamefont {Reimers}}]{Sajid_2020_rev}%
  \BibitemOpen
  \bibfield  {author} {\bibinfo {author} {\bibfnamefont {A.}~\bibnamefont
  {Sajid}}, \bibinfo {author} {\bibfnamefont {M.~J.}\ \bibnamefont {Ford}}, \
  and\ \bibinfo {author} {\bibfnamefont {J.~R.}\ \bibnamefont {Reimers}},\
  }\href {\doibase 10.1088/1361-6633/ab6310} {\bibfield  {journal} {\bibinfo
  {journal} {Reports on Progress in Physics}\ }\textbf {\bibinfo {volume}
  {83}},\ \bibinfo {pages} {044501} (\bibinfo {year}
  {2020}{\natexlab{a}})}\BibitemShut {NoStop}%
\bibitem [{\citenamefont {Chejanovsky}\ \emph {et~al.}(2021)\citenamefont
  {Chejanovsky}, \citenamefont {Mukherjee}, \citenamefont {Geng}, \citenamefont
  {Chen}, \citenamefont {Kim}, \citenamefont {Denisenko}, \citenamefont
  {Finkler}, \citenamefont {Taniguchi}, \citenamefont {Watanabe}, \citenamefont
  {Dasari}, \citenamefont {Auburger}, \citenamefont {Gali}, \citenamefont
  {Smet},\ and\ \citenamefont {Wrachtrup}}]{Chejanovsky_2021}%
  \BibitemOpen
  \bibfield  {author} {\bibinfo {author} {\bibfnamefont {N.}~\bibnamefont
  {Chejanovsky}}, \bibinfo {author} {\bibfnamefont {A.}~\bibnamefont
  {Mukherjee}}, \bibinfo {author} {\bibfnamefont {J.}~\bibnamefont {Geng}},
  \bibinfo {author} {\bibfnamefont {Y.-C.}\ \bibnamefont {Chen}}, \bibinfo
  {author} {\bibfnamefont {Y.}~\bibnamefont {Kim}}, \bibinfo {author}
  {\bibfnamefont {A.}~\bibnamefont {Denisenko}}, \bibinfo {author}
  {\bibfnamefont {A.}~\bibnamefont {Finkler}}, \bibinfo {author} {\bibfnamefont
  {T.}~\bibnamefont {Taniguchi}}, \bibinfo {author} {\bibfnamefont
  {K.}~\bibnamefont {Watanabe}}, \bibinfo {author} {\bibfnamefont {D.~B.~R.}\
  \bibnamefont {Dasari}}, \bibinfo {author} {\bibfnamefont {P.}~\bibnamefont
  {Auburger}}, \bibinfo {author} {\bibfnamefont {A.}~\bibnamefont {Gali}},
  \bibinfo {author} {\bibfnamefont {J.~H.}\ \bibnamefont {Smet}}, \ and\
  \bibinfo {author} {\bibfnamefont {J.}~\bibnamefont {Wrachtrup}},\ }\href
  {\doibase 10.1038/s41563-021-00979-4} {\bibfield  {journal} {\bibinfo
  {journal} {Nature Materials}\ }\textbf {\bibinfo {volume} {20}},\ \bibinfo
  {pages} {1079} (\bibinfo {year} {2021})}\BibitemShut {NoStop}%
\bibitem [{\citenamefont {Stern}\ \emph {et~al.}(2022)\citenamefont {Stern},
  \citenamefont {Gu}, \citenamefont {Jarman}, \citenamefont {Eizagirre~Barker},
  \citenamefont {Mendelson}, \citenamefont {Chugh}, \citenamefont {Schott},
  \citenamefont {Tan}, \citenamefont {Sirringhaus}, \citenamefont
  {Aharonovich},\ and\ \citenamefont {Atat{\"u}re}}]{Stern_2022}%
  \BibitemOpen
  \bibfield  {author} {\bibinfo {author} {\bibfnamefont {H.~L.}\ \bibnamefont
  {Stern}}, \bibinfo {author} {\bibfnamefont {Q.}~\bibnamefont {Gu}}, \bibinfo
  {author} {\bibfnamefont {J.}~\bibnamefont {Jarman}}, \bibinfo {author}
  {\bibfnamefont {S.}~\bibnamefont {Eizagirre~Barker}}, \bibinfo {author}
  {\bibfnamefont {N.}~\bibnamefont {Mendelson}}, \bibinfo {author}
  {\bibfnamefont {D.}~\bibnamefont {Chugh}}, \bibinfo {author} {\bibfnamefont
  {S.}~\bibnamefont {Schott}}, \bibinfo {author} {\bibfnamefont {H.~H.}\
  \bibnamefont {Tan}}, \bibinfo {author} {\bibfnamefont {H.}~\bibnamefont
  {Sirringhaus}}, \bibinfo {author} {\bibfnamefont {I.}~\bibnamefont
  {Aharonovich}}, \ and\ \bibinfo {author} {\bibfnamefont {M.}~\bibnamefont
  {Atat{\"u}re}},\ }\href {\doibase 10.1038/s41467-022-28169-z} {\bibfield
  {journal} {\bibinfo  {journal} {Nature Communications}\ }\textbf {\bibinfo
  {volume} {13}},\ \bibinfo {pages} {618} (\bibinfo {year} {2022})}\BibitemShut
  {NoStop}%
\bibitem [{\citenamefont {Sajid}\ \emph {et~al.}(2018)\citenamefont {Sajid},
  \citenamefont {Reimers},\ and\ \citenamefont {Ford}}]{Sajid_2018}%
  \BibitemOpen
  \bibfield  {author} {\bibinfo {author} {\bibfnamefont {A.}~\bibnamefont
  {Sajid}}, \bibinfo {author} {\bibfnamefont {J.~R.}\ \bibnamefont {Reimers}},
  \ and\ \bibinfo {author} {\bibfnamefont {M.~J.}\ \bibnamefont {Ford}},\
  }\href {\doibase 10.1103/PhysRevB.97.064101} {\bibfield  {journal} {\bibinfo
  {journal} {Phys. Rev. B}\ }\textbf {\bibinfo {volume} {97}},\ \bibinfo
  {pages} {064101} (\bibinfo {year} {2018})}\BibitemShut {NoStop}%
\bibitem [{\citenamefont {Weston}\ \emph {et~al.}(2018)\citenamefont {Weston},
  \citenamefont {Wickramaratne}, \citenamefont {Mackoit}, \citenamefont
  {Alkauskas},\ and\ \citenamefont {Van~de Walle}}]{Weston_2018}%
  \BibitemOpen
  \bibfield  {author} {\bibinfo {author} {\bibfnamefont {L.}~\bibnamefont
  {Weston}}, \bibinfo {author} {\bibfnamefont {D.}~\bibnamefont
  {Wickramaratne}}, \bibinfo {author} {\bibfnamefont {M.}~\bibnamefont
  {Mackoit}}, \bibinfo {author} {\bibfnamefont {A.}~\bibnamefont {Alkauskas}},
  \ and\ \bibinfo {author} {\bibfnamefont {C.~G.}\ \bibnamefont {Van~de
  Walle}},\ }\href {\doibase 10.1103/PhysRevB.97.214104} {\bibfield  {journal}
  {\bibinfo  {journal} {Phys. Rev. B}\ }\textbf {\bibinfo {volume} {97}},\
  \bibinfo {pages} {214104} (\bibinfo {year} {2018})}\BibitemShut {NoStop}%
\bibitem [{\citenamefont {Sajid}\ \emph
  {et~al.}(2020{\natexlab{b}})\citenamefont {Sajid}, \citenamefont {Reimers},
  \citenamefont {Kobayashi},\ and\ \citenamefont {Ford}}]{Sajid_2020}%
  \BibitemOpen
  \bibfield  {author} {\bibinfo {author} {\bibfnamefont {A.}~\bibnamefont
  {Sajid}}, \bibinfo {author} {\bibfnamefont {J.~R.}\ \bibnamefont {Reimers}},
  \bibinfo {author} {\bibfnamefont {R.}~\bibnamefont {Kobayashi}}, \ and\
  \bibinfo {author} {\bibfnamefont {M.~J.}\ \bibnamefont {Ford}},\ }\href
  {\doibase 10.1103/PhysRevB.102.144104} {\bibfield  {journal} {\bibinfo
  {journal} {Phys. Rev. B}\ }\textbf {\bibinfo {volume} {102}},\ \bibinfo
  {pages} {144104} (\bibinfo {year} {2020}{\natexlab{b}})}\BibitemShut
  {NoStop}%
\bibitem [{\citenamefont {Auburger}\ and\ \citenamefont
  {Gali}(2021)}]{Auburger_2021}%
  \BibitemOpen
  \bibfield  {author} {\bibinfo {author} {\bibfnamefont {P.}~\bibnamefont
  {Auburger}}\ and\ \bibinfo {author} {\bibfnamefont {A.}~\bibnamefont
  {Gali}},\ }\href {\doibase 10.1103/PhysRevB.104.075410} {\bibfield  {journal}
  {\bibinfo  {journal} {Phys. Rev. B}\ }\textbf {\bibinfo {volume} {104}},\
  \bibinfo {pages} {075410} (\bibinfo {year} {2021})}\BibitemShut {NoStop}%
\bibitem [{\citenamefont {Gottscholl}\ \emph {et~al.}(2020)\citenamefont
  {Gottscholl}, \citenamefont {Kianinia}, \citenamefont {Soltamov},
  \citenamefont {Orlinskii}, \citenamefont {Mamin}, \citenamefont {Bradac},
  \citenamefont {Kasper}, \citenamefont {Krambrock}, \citenamefont {Sperlich},
  \citenamefont {Toth}, \citenamefont {Aharonovich},\ and\ \citenamefont
  {Dyakonov}}]{Gottscholl2020}%
  \BibitemOpen
  \bibfield  {author} {\bibinfo {author} {\bibfnamefont {A.}~\bibnamefont
  {Gottscholl}}, \bibinfo {author} {\bibfnamefont {M.}~\bibnamefont
  {Kianinia}}, \bibinfo {author} {\bibfnamefont {V.}~\bibnamefont {Soltamov}},
  \bibinfo {author} {\bibfnamefont {S.}~\bibnamefont {Orlinskii}}, \bibinfo
  {author} {\bibfnamefont {G.}~\bibnamefont {Mamin}}, \bibinfo {author}
  {\bibfnamefont {C.}~\bibnamefont {Bradac}}, \bibinfo {author} {\bibfnamefont
  {C.}~\bibnamefont {Kasper}}, \bibinfo {author} {\bibfnamefont
  {K.}~\bibnamefont {Krambrock}}, \bibinfo {author} {\bibfnamefont
  {A.}~\bibnamefont {Sperlich}}, \bibinfo {author} {\bibfnamefont
  {M.}~\bibnamefont {Toth}}, \bibinfo {author} {\bibfnamefont {I.}~\bibnamefont
  {Aharonovich}}, \ and\ \bibinfo {author} {\bibfnamefont {V.}~\bibnamefont
  {Dyakonov}},\ }\href {\doibase 10.1038/s41563-020-0619-6} {\bibfield
  {journal} {\bibinfo  {journal} {Nature Materials}\ }\textbf {\bibinfo
  {volume} {19}},\ \bibinfo {pages} {540} (\bibinfo {year} {2020})}\BibitemShut
  {NoStop}%
\bibitem [{\citenamefont {Abdi}\ \emph {et~al.}(2018)\citenamefont {Abdi},
  \citenamefont {Chou}, \citenamefont {Gali},\ and\ \citenamefont
  {Plenio}}]{Abdi_2018}%
  \BibitemOpen
  \bibfield  {author} {\bibinfo {author} {\bibfnamefont {M.}~\bibnamefont
  {Abdi}}, \bibinfo {author} {\bibfnamefont {J.-P.}\ \bibnamefont {Chou}},
  \bibinfo {author} {\bibfnamefont {A.}~\bibnamefont {Gali}}, \ and\ \bibinfo
  {author} {\bibfnamefont {M.~B.}\ \bibnamefont {Plenio}},\ }\href {\doibase
  10.1021/acsphotonics.7b01442} {\bibfield  {journal} {\bibinfo  {journal} {ACS
  Photonics}\ }\textbf {\bibinfo {volume} {5}},\ \bibinfo {pages} {1967}
  (\bibinfo {year} {2018})}\BibitemShut {NoStop}%
\bibitem [{\citenamefont {Iv{\'a}dy}\ \emph {et~al.}(2020)\citenamefont
  {Iv{\'a}dy}, \citenamefont {Barcza}, \citenamefont {Thiering}, \citenamefont
  {Li}, \citenamefont {Hamdi}, \citenamefont {Chou}, \citenamefont {Legeza},\
  and\ \citenamefont {Gali}}]{Ivady_2020}%
  \BibitemOpen
  \bibfield  {author} {\bibinfo {author} {\bibfnamefont {V.}~\bibnamefont
  {Iv{\'a}dy}}, \bibinfo {author} {\bibfnamefont {G.}~\bibnamefont {Barcza}},
  \bibinfo {author} {\bibfnamefont {G.}~\bibnamefont {Thiering}}, \bibinfo
  {author} {\bibfnamefont {S.}~\bibnamefont {Li}}, \bibinfo {author}
  {\bibfnamefont {H.}~\bibnamefont {Hamdi}}, \bibinfo {author} {\bibfnamefont
  {J.-P.}\ \bibnamefont {Chou}}, \bibinfo {author} {\bibfnamefont
  {{\"O}.}~\bibnamefont {Legeza}}, \ and\ \bibinfo {author} {\bibfnamefont
  {A.}~\bibnamefont {Gali}},\ }\href {\doibase 10.1038/s41524-020-0305-x}
  {\bibfield  {journal} {\bibinfo  {journal} {npj Computational Materials}\
  }\textbf {\bibinfo {volume} {6}},\ \bibinfo {pages} {41} (\bibinfo {year}
  {2020})}\BibitemShut {NoStop}%
\bibitem [{\citenamefont {Haykal}\ \emph {et~al.}(2022)\citenamefont {Haykal},
  \citenamefont {Tanos}, \citenamefont {Minotto}, \citenamefont {Durand},
  \citenamefont {Fabre}, \citenamefont {Li}, \citenamefont {Edgar},
  \citenamefont {Iv{\'a}dy}, \citenamefont {Gali}, \citenamefont {Michel},
  \citenamefont {Dr{\'e}au}, \citenamefont {Gil}, \citenamefont {Cassabois},\
  and\ \citenamefont {Jacques}}]{haykal2022}%
  \BibitemOpen
  \bibfield  {author} {\bibinfo {author} {\bibfnamefont {A.}~\bibnamefont
  {Haykal}}, \bibinfo {author} {\bibfnamefont {R.}~\bibnamefont {Tanos}},
  \bibinfo {author} {\bibfnamefont {N.}~\bibnamefont {Minotto}}, \bibinfo
  {author} {\bibfnamefont {A.}~\bibnamefont {Durand}}, \bibinfo {author}
  {\bibfnamefont {F.}~\bibnamefont {Fabre}}, \bibinfo {author} {\bibfnamefont
  {J.}~\bibnamefont {Li}}, \bibinfo {author} {\bibfnamefont {J.~H.}\
  \bibnamefont {Edgar}}, \bibinfo {author} {\bibfnamefont {V.}~\bibnamefont
  {Iv{\'a}dy}}, \bibinfo {author} {\bibfnamefont {A.}~\bibnamefont {Gali}},
  \bibinfo {author} {\bibfnamefont {T.}~\bibnamefont {Michel}}, \bibinfo
  {author} {\bibfnamefont {A.}~\bibnamefont {Dr{\'e}au}}, \bibinfo {author}
  {\bibfnamefont {B.}~\bibnamefont {Gil}}, \bibinfo {author} {\bibfnamefont
  {G.}~\bibnamefont {Cassabois}}, \ and\ \bibinfo {author} {\bibfnamefont
  {V.}~\bibnamefont {Jacques}},\ }\href
  {https://doi.org/10.1038/s41467-022-31743-0} {\bibfield  {journal} {\bibinfo
  {journal} {Nature Communications}\ }\textbf {\bibinfo {volume} {13}},\
  \bibinfo {pages} {4347} (\bibinfo {year} {2022})}\BibitemShut {NoStop}%
\bibitem [{\citenamefont {Li}\ \emph {et~al.}(2021)\citenamefont {Li},
  \citenamefont {Glaser}, \citenamefont {Elias}, \citenamefont {Ye},
  \citenamefont {Evans}, \citenamefont {Xue}, \citenamefont {Liu},
  \citenamefont {Cassabois}, \citenamefont {Gil}, \citenamefont {Valvin},
  \citenamefont {Pelini}, \citenamefont {Yeats}, \citenamefont {He},
  \citenamefont {Liu},\ and\ \citenamefont {Edgar}}]{Li_2021}%
  \BibitemOpen
  \bibfield  {author} {\bibinfo {author} {\bibfnamefont {J.}~\bibnamefont
  {Li}}, \bibinfo {author} {\bibfnamefont {E.~R.}\ \bibnamefont {Glaser}},
  \bibinfo {author} {\bibfnamefont {C.}~\bibnamefont {Elias}}, \bibinfo
  {author} {\bibfnamefont {G.}~\bibnamefont {Ye}}, \bibinfo {author}
  {\bibfnamefont {D.}~\bibnamefont {Evans}}, \bibinfo {author} {\bibfnamefont
  {L.}~\bibnamefont {Xue}}, \bibinfo {author} {\bibfnamefont {S.}~\bibnamefont
  {Liu}}, \bibinfo {author} {\bibfnamefont {G.}~\bibnamefont {Cassabois}},
  \bibinfo {author} {\bibfnamefont {B.}~\bibnamefont {Gil}}, \bibinfo {author}
  {\bibfnamefont {P.}~\bibnamefont {Valvin}}, \bibinfo {author} {\bibfnamefont
  {T.}~\bibnamefont {Pelini}}, \bibinfo {author} {\bibfnamefont {A.~L.}\
  \bibnamefont {Yeats}}, \bibinfo {author} {\bibfnamefont {R.}~\bibnamefont
  {He}}, \bibinfo {author} {\bibfnamefont {B.}~\bibnamefont {Liu}}, \ and\
  \bibinfo {author} {\bibfnamefont {J.~H.}\ \bibnamefont {Edgar}},\ }\href
  {\doibase 10.1021/acs.chemmater.1c02849} {\bibfield  {journal} {\bibinfo
  {journal} {Chemistry of Materials}\ }\textbf {\bibinfo {volume} {33}},\
  \bibinfo {pages} {9231} (\bibinfo {year} {2021})}\BibitemShut {NoStop}%
\bibitem [{\citenamefont {Murzakhanov}\ \emph {et~al.}(2021)\citenamefont
  {Murzakhanov}, \citenamefont {Yavkin}, \citenamefont {Mamin}, \citenamefont
  {Orlinskii}, \citenamefont {Mumdzhi}, \citenamefont {Gracheva}, \citenamefont
  {Gabbasov}, \citenamefont {Smirnov}, \citenamefont {Davydov},\ and\
  \citenamefont {Soltamov}}]{Murzakhanov_2021}%
  \BibitemOpen
  \bibfield  {author} {\bibinfo {author} {\bibfnamefont {F.~F.}\ \bibnamefont
  {Murzakhanov}}, \bibinfo {author} {\bibfnamefont {B.~V.}\ \bibnamefont
  {Yavkin}}, \bibinfo {author} {\bibfnamefont {G.~V.}\ \bibnamefont {Mamin}},
  \bibinfo {author} {\bibfnamefont {S.~B.}\ \bibnamefont {Orlinskii}}, \bibinfo
  {author} {\bibfnamefont {I.~E.}\ \bibnamefont {Mumdzhi}}, \bibinfo {author}
  {\bibfnamefont {I.~N.}\ \bibnamefont {Gracheva}}, \bibinfo {author}
  {\bibfnamefont {B.~F.}\ \bibnamefont {Gabbasov}}, \bibinfo {author}
  {\bibfnamefont {A.~N.}\ \bibnamefont {Smirnov}}, \bibinfo {author}
  {\bibfnamefont {V.~Y.}\ \bibnamefont {Davydov}}, \ and\ \bibinfo {author}
  {\bibfnamefont {V.~A.}\ \bibnamefont {Soltamov}},\ }\href {\doibase
  10.3390/nano11061373} {\bibfield  {journal} {\bibinfo  {journal}
  {Nanomaterials}\ }\textbf {\bibinfo {volume} {11}} (\bibinfo {year} {2021}),\
  10.3390/nano11061373}\BibitemShut {NoStop}%
\bibitem [{\citenamefont {Kianinia}\ \emph {et~al.}(2020)\citenamefont
  {Kianinia}, \citenamefont {White}, \citenamefont {Fröch}, \citenamefont
  {Bradac},\ and\ \citenamefont {Aharonovich}}]{Kianinia_2020}%
  \BibitemOpen
  \bibfield  {author} {\bibinfo {author} {\bibfnamefont {M.}~\bibnamefont
  {Kianinia}}, \bibinfo {author} {\bibfnamefont {S.}~\bibnamefont {White}},
  \bibinfo {author} {\bibfnamefont {J.~E.}\ \bibnamefont {Fröch}}, \bibinfo
  {author} {\bibfnamefont {C.}~\bibnamefont {Bradac}}, \ and\ \bibinfo {author}
  {\bibfnamefont {I.}~\bibnamefont {Aharonovich}},\ }\href {\doibase
  10.1021/acsphotonics.0c00614} {\bibfield  {journal} {\bibinfo  {journal} {ACS
  Photonics}\ }\textbf {\bibinfo {volume} {7}},\ \bibinfo {pages} {2147}
  (\bibinfo {year} {2020})}\BibitemShut {NoStop}%
\bibitem [{\citenamefont {Guo}\ \emph {et~al.}(2022)\citenamefont {Guo},
  \citenamefont {Liu}, \citenamefont {Li}, \citenamefont {Yang}, \citenamefont
  {Yu}, \citenamefont {Meng}, \citenamefont {Wang}, \citenamefont {Zeng},
  \citenamefont {Yan}, \citenamefont {Li}, \citenamefont {Wang}, \citenamefont
  {Xu}, \citenamefont {Wang}, \citenamefont {Tang}, \citenamefont {Li},\ and\
  \citenamefont {Guo}}]{Guo_2022}%
  \BibitemOpen
  \bibfield  {author} {\bibinfo {author} {\bibfnamefont {N.-J.}\ \bibnamefont
  {Guo}}, \bibinfo {author} {\bibfnamefont {W.}~\bibnamefont {Liu}}, \bibinfo
  {author} {\bibfnamefont {Z.-P.}\ \bibnamefont {Li}}, \bibinfo {author}
  {\bibfnamefont {Y.-Z.}\ \bibnamefont {Yang}}, \bibinfo {author}
  {\bibfnamefont {S.}~\bibnamefont {Yu}}, \bibinfo {author} {\bibfnamefont
  {Y.}~\bibnamefont {Meng}}, \bibinfo {author} {\bibfnamefont {Z.-A.}\
  \bibnamefont {Wang}}, \bibinfo {author} {\bibfnamefont {X.-D.}\ \bibnamefont
  {Zeng}}, \bibinfo {author} {\bibfnamefont {F.-F.}\ \bibnamefont {Yan}},
  \bibinfo {author} {\bibfnamefont {Q.}~\bibnamefont {Li}}, \bibinfo {author}
  {\bibfnamefont {J.-F.}\ \bibnamefont {Wang}}, \bibinfo {author}
  {\bibfnamefont {J.-S.}\ \bibnamefont {Xu}}, \bibinfo {author} {\bibfnamefont
  {Y.-T.}\ \bibnamefont {Wang}}, \bibinfo {author} {\bibfnamefont {J.-S.}\
  \bibnamefont {Tang}}, \bibinfo {author} {\bibfnamefont {C.-F.}\ \bibnamefont
  {Li}}, \ and\ \bibinfo {author} {\bibfnamefont {G.-C.}\ \bibnamefont {Guo}},\
  }\href {\doibase 10.1021/acsomega.1c04564} {\bibfield  {journal} {\bibinfo
  {journal} {ACS Omega}\ }\textbf {\bibinfo {volume} {7}},\ \bibinfo {pages}
  {1733} (\bibinfo {year} {2022})}\BibitemShut {NoStop}%
\bibitem [{\citenamefont {Gao}\ \emph {et~al.}(2021{\natexlab{a}})\citenamefont
  {Gao}, \citenamefont {Pandey}, \citenamefont {Kianinia}, \citenamefont {Ahn},
  \citenamefont {Ju}, \citenamefont {Aharonovich}, \citenamefont {Shivaram},\
  and\ \citenamefont {Li}}]{Gao2021}%
  \BibitemOpen
  \bibfield  {author} {\bibinfo {author} {\bibfnamefont {X.}~\bibnamefont
  {Gao}}, \bibinfo {author} {\bibfnamefont {S.}~\bibnamefont {Pandey}},
  \bibinfo {author} {\bibfnamefont {M.}~\bibnamefont {Kianinia}}, \bibinfo
  {author} {\bibfnamefont {J.}~\bibnamefont {Ahn}}, \bibinfo {author}
  {\bibfnamefont {P.}~\bibnamefont {Ju}}, \bibinfo {author} {\bibfnamefont
  {I.}~\bibnamefont {Aharonovich}}, \bibinfo {author} {\bibfnamefont
  {N.}~\bibnamefont {Shivaram}}, \ and\ \bibinfo {author} {\bibfnamefont
  {T.}~\bibnamefont {Li}},\ }\href {\doibase 10.1021/acsphotonics.0c01847}
  {\bibfield  {journal} {\bibinfo  {journal} {ACS Photonics}\ }\textbf
  {\bibinfo {volume} {8}},\ \bibinfo {pages} {994} (\bibinfo {year}
  {2021}{\natexlab{a}})}\BibitemShut {NoStop}%
\bibitem [{\citenamefont {Gottscholl}\ \emph {et~al.}(2021)\citenamefont
  {Gottscholl}, \citenamefont {Diez}, \citenamefont {Soltamov}, \citenamefont
  {Kasper}, \citenamefont {Krau{\ss}e}, \citenamefont {Sperlich}, \citenamefont
  {Kianinia}, \citenamefont {Bradac}, \citenamefont {Aharonovich},\ and\
  \citenamefont {Dyakonov}}]{Gottscholl2021}%
  \BibitemOpen
  \bibfield  {author} {\bibinfo {author} {\bibfnamefont {A.}~\bibnamefont
  {Gottscholl}}, \bibinfo {author} {\bibfnamefont {M.}~\bibnamefont {Diez}},
  \bibinfo {author} {\bibfnamefont {V.}~\bibnamefont {Soltamov}}, \bibinfo
  {author} {\bibfnamefont {C.}~\bibnamefont {Kasper}}, \bibinfo {author}
  {\bibfnamefont {D.}~\bibnamefont {Krau{\ss}e}}, \bibinfo {author}
  {\bibfnamefont {A.}~\bibnamefont {Sperlich}}, \bibinfo {author}
  {\bibfnamefont {M.}~\bibnamefont {Kianinia}}, \bibinfo {author}
  {\bibfnamefont {C.}~\bibnamefont {Bradac}}, \bibinfo {author} {\bibfnamefont
  {I.}~\bibnamefont {Aharonovich}}, \ and\ \bibinfo {author} {\bibfnamefont
  {V.}~\bibnamefont {Dyakonov}},\ }\href {\doibase 10.1038/s41467-021-24725-1}
  {\bibfield  {journal} {\bibinfo  {journal} {Nature Communications}\ }\textbf
  {\bibinfo {volume} {12}},\ \bibinfo {pages} {4480} (\bibinfo {year}
  {2021})}\BibitemShut {NoStop}%
\bibitem [{\citenamefont {Gao}\ \emph {et~al.}(2021{\natexlab{b}})\citenamefont
  {Gao}, \citenamefont {Jiang}, \citenamefont {Llacsahuanga~Allcca},
  \citenamefont {Shen}, \citenamefont {Sadi}, \citenamefont {Solanki},
  \citenamefont {Ju}, \citenamefont {Xu}, \citenamefont {Upadhyaya},
  \citenamefont {Chen}, \citenamefont {Bhave},\ and\ \citenamefont
  {Li}}]{Gao_2021}%
  \BibitemOpen
  \bibfield  {author} {\bibinfo {author} {\bibfnamefont {X.}~\bibnamefont
  {Gao}}, \bibinfo {author} {\bibfnamefont {B.}~\bibnamefont {Jiang}}, \bibinfo
  {author} {\bibfnamefont {A.~E.}\ \bibnamefont {Llacsahuanga~Allcca}},
  \bibinfo {author} {\bibfnamefont {K.}~\bibnamefont {Shen}}, \bibinfo {author}
  {\bibfnamefont {M.~A.}\ \bibnamefont {Sadi}}, \bibinfo {author}
  {\bibfnamefont {A.~B.}\ \bibnamefont {Solanki}}, \bibinfo {author}
  {\bibfnamefont {P.}~\bibnamefont {Ju}}, \bibinfo {author} {\bibfnamefont
  {Z.}~\bibnamefont {Xu}}, \bibinfo {author} {\bibfnamefont {P.}~\bibnamefont
  {Upadhyaya}}, \bibinfo {author} {\bibfnamefont {Y.~P.}\ \bibnamefont {Chen}},
  \bibinfo {author} {\bibfnamefont {S.~A.}\ \bibnamefont {Bhave}}, \ and\
  \bibinfo {author} {\bibfnamefont {T.}~\bibnamefont {Li}},\ }\href
  {https://doi.org/10.1021/acs.nanolett.1c02495} {\bibfield  {journal}
  {\bibinfo  {journal} {Nano Letters}\ }\textbf {\bibinfo {volume} {21}},\
  \bibinfo {pages} {7708} (\bibinfo {year} {2021}{\natexlab{b}})}\BibitemShut
  {NoStop}%
\bibitem [{\citenamefont {Huang}\ \emph {et~al.}(2022)\citenamefont {Huang},
  \citenamefont {Zhou}, \citenamefont {Chen}, \citenamefont {Lu}, \citenamefont
  {McLaughlin}, \citenamefont {Li}, \citenamefont {Alghamdi}, \citenamefont
  {Djugba}, \citenamefont {Shi}, \citenamefont {Wang},\ and\ \citenamefont
  {Du}}]{Huang_2022}%
  \BibitemOpen
  \bibfield  {author} {\bibinfo {author} {\bibfnamefont {M.}~\bibnamefont
  {Huang}}, \bibinfo {author} {\bibfnamefont {J.}~\bibnamefont {Zhou}},
  \bibinfo {author} {\bibfnamefont {D.}~\bibnamefont {Chen}}, \bibinfo {author}
  {\bibfnamefont {H.}~\bibnamefont {Lu}}, \bibinfo {author} {\bibfnamefont
  {N.~J.}\ \bibnamefont {McLaughlin}}, \bibinfo {author} {\bibfnamefont
  {S.}~\bibnamefont {Li}}, \bibinfo {author} {\bibfnamefont {M.}~\bibnamefont
  {Alghamdi}}, \bibinfo {author} {\bibfnamefont {D.}~\bibnamefont {Djugba}},
  \bibinfo {author} {\bibfnamefont {J.}~\bibnamefont {Shi}}, \bibinfo {author}
  {\bibfnamefont {H.}~\bibnamefont {Wang}}, \ and\ \bibinfo {author}
  {\bibfnamefont {C.~R.}\ \bibnamefont {Du}},\ }\href {\doibase
  10.1038/s41467-022-33016-2} {\bibfield  {journal} {\bibinfo  {journal}
  {Nature Communications}\ }\textbf {\bibinfo {volume} {13}},\ \bibinfo {pages}
  {5369} (\bibinfo {year} {2022})}\BibitemShut {NoStop}%
\bibitem [{\citenamefont {Kumar}\ \emph {et~al.}(2022)\citenamefont {Kumar},
  \citenamefont {Fabre}, \citenamefont {Durand}, \citenamefont {Clua-Provost},
  \citenamefont {Li}, \citenamefont {Edgar}, \citenamefont {Rougemaille},
  \citenamefont {Coraux}, \citenamefont {Marie}, \citenamefont {Renucci},
  \citenamefont {Robert}, \citenamefont {Robert-Philip}, \citenamefont {Gil},
  \citenamefont {Cassabois}, \citenamefont {Finco},\ and\ \citenamefont
  {Jacques}}]{Kumar_2022}%
  \BibitemOpen
  \bibfield  {author} {\bibinfo {author} {\bibfnamefont {P.}~\bibnamefont
  {Kumar}}, \bibinfo {author} {\bibfnamefont {F.}~\bibnamefont {Fabre}},
  \bibinfo {author} {\bibfnamefont {A.}~\bibnamefont {Durand}}, \bibinfo
  {author} {\bibfnamefont {T.}~\bibnamefont {Clua-Provost}}, \bibinfo {author}
  {\bibfnamefont {J.}~\bibnamefont {Li}}, \bibinfo {author} {\bibfnamefont
  {J.}~\bibnamefont {Edgar}}, \bibinfo {author} {\bibfnamefont
  {N.}~\bibnamefont {Rougemaille}}, \bibinfo {author} {\bibfnamefont
  {J.}~\bibnamefont {Coraux}}, \bibinfo {author} {\bibfnamefont
  {X.}~\bibnamefont {Marie}}, \bibinfo {author} {\bibfnamefont
  {P.}~\bibnamefont {Renucci}}, \bibinfo {author} {\bibfnamefont
  {C.}~\bibnamefont {Robert}}, \bibinfo {author} {\bibfnamefont
  {I.}~\bibnamefont {Robert-Philip}}, \bibinfo {author} {\bibfnamefont
  {B.}~\bibnamefont {Gil}}, \bibinfo {author} {\bibfnamefont {G.}~\bibnamefont
  {Cassabois}}, \bibinfo {author} {\bibfnamefont {A.}~\bibnamefont {Finco}}, \
  and\ \bibinfo {author} {\bibfnamefont {V.}~\bibnamefont {Jacques}},\ }\href
  {\doibase 10.1103/PhysRevApplied.18.L061002} {\bibfield  {journal} {\bibinfo
  {journal} {Phys. Rev. Appl.}\ }\textbf {\bibinfo {volume} {18}},\ \bibinfo
  {pages} {L061002} (\bibinfo {year} {2022})}\BibitemShut {NoStop}%
\bibitem [{\citenamefont {Healey}\ \emph {et~al.}(2023)\citenamefont {Healey},
  \citenamefont {Scholten}, \citenamefont {Yang}, \citenamefont {Scott},
  \citenamefont {Abrahams}, \citenamefont {Robertson}, \citenamefont {Hou},
  \citenamefont {Guo}, \citenamefont {Rahman}, \citenamefont {Lu},
  \citenamefont {Kianinia}, \citenamefont {Aharonovich},\ and\ \citenamefont
  {Tetienne}}]{Healey_2023}%
  \BibitemOpen
  \bibfield  {author} {\bibinfo {author} {\bibfnamefont {A.~J.}\ \bibnamefont
  {Healey}}, \bibinfo {author} {\bibfnamefont {S.~C.}\ \bibnamefont
  {Scholten}}, \bibinfo {author} {\bibfnamefont {T.}~\bibnamefont {Yang}},
  \bibinfo {author} {\bibfnamefont {J.~A.}\ \bibnamefont {Scott}}, \bibinfo
  {author} {\bibfnamefont {G.~J.}\ \bibnamefont {Abrahams}}, \bibinfo {author}
  {\bibfnamefont {I.~O.}\ \bibnamefont {Robertson}}, \bibinfo {author}
  {\bibfnamefont {X.~F.}\ \bibnamefont {Hou}}, \bibinfo {author} {\bibfnamefont
  {Y.~F.}\ \bibnamefont {Guo}}, \bibinfo {author} {\bibfnamefont
  {S.}~\bibnamefont {Rahman}}, \bibinfo {author} {\bibfnamefont
  {Y.}~\bibnamefont {Lu}}, \bibinfo {author} {\bibfnamefont {M.}~\bibnamefont
  {Kianinia}}, \bibinfo {author} {\bibfnamefont {I.}~\bibnamefont
  {Aharonovich}}, \ and\ \bibinfo {author} {\bibfnamefont {J.-P.}\ \bibnamefont
  {Tetienne}},\ }\href {\doibase 10.1038/s41567-022-01815-5} {\bibfield
  {journal} {\bibinfo  {journal} {Nature Physics}\ }\textbf {\bibinfo {volume}
  {19}},\ \bibinfo {pages} {87} (\bibinfo {year} {2023})}\BibitemShut {NoStop}%
\bibitem [{\citenamefont {Yang}\ \emph {et~al.}(2022)\citenamefont {Yang},
  \citenamefont {Mendelson}, \citenamefont {Li}, \citenamefont {Gottscholl},
  \citenamefont {Scott}, \citenamefont {Kianinia}, \citenamefont {Dyakonov},
  \citenamefont {Toth},\ and\ \citenamefont {Aharonovich}}]{Yang_2022}%
  \BibitemOpen
  \bibfield  {author} {\bibinfo {author} {\bibfnamefont {T.}~\bibnamefont
  {Yang}}, \bibinfo {author} {\bibfnamefont {N.}~\bibnamefont {Mendelson}},
  \bibinfo {author} {\bibfnamefont {C.}~\bibnamefont {Li}}, \bibinfo {author}
  {\bibfnamefont {A.}~\bibnamefont {Gottscholl}}, \bibinfo {author}
  {\bibfnamefont {J.}~\bibnamefont {Scott}}, \bibinfo {author} {\bibfnamefont
  {M.}~\bibnamefont {Kianinia}}, \bibinfo {author} {\bibfnamefont
  {V.}~\bibnamefont {Dyakonov}}, \bibinfo {author} {\bibfnamefont
  {M.}~\bibnamefont {Toth}}, \ and\ \bibinfo {author} {\bibfnamefont
  {I.}~\bibnamefont {Aharonovich}},\ }\href {\doibase 10.1039/D1NR07919K}
  {\bibfield  {journal} {\bibinfo  {journal} {Nanoscale}\ }\textbf {\bibinfo
  {volume} {14}},\ \bibinfo {pages} {5239} (\bibinfo {year}
  {2022})}\BibitemShut {NoStop}%
\bibitem [{\citenamefont {Lyu}\ \emph {et~al.}(2022)\citenamefont {Lyu},
  \citenamefont {Tan}, \citenamefont {Wu}, \citenamefont {Zhang}, \citenamefont
  {Zhang}, \citenamefont {Mu}, \citenamefont {Z{\'u}{\~n}iga-P{\'e}rez},
  \citenamefont {Cai},\ and\ \citenamefont {Gao}}]{GaoStrain2022}%
  \BibitemOpen
  \bibfield  {author} {\bibinfo {author} {\bibfnamefont {X.}~\bibnamefont
  {Lyu}}, \bibinfo {author} {\bibfnamefont {Q.}~\bibnamefont {Tan}}, \bibinfo
  {author} {\bibfnamefont {L.}~\bibnamefont {Wu}}, \bibinfo {author}
  {\bibfnamefont {C.}~\bibnamefont {Zhang}}, \bibinfo {author} {\bibfnamefont
  {Z.}~\bibnamefont {Zhang}}, \bibinfo {author} {\bibfnamefont
  {Z.}~\bibnamefont {Mu}}, \bibinfo {author} {\bibfnamefont {J.}~\bibnamefont
  {Z{\'u}{\~n}iga-P{\'e}rez}}, \bibinfo {author} {\bibfnamefont
  {H.}~\bibnamefont {Cai}}, \ and\ \bibinfo {author} {\bibfnamefont
  {W.}~\bibnamefont {Gao}},\ }\href
  {https://doi.org/10.1021/acs.nanolett.2c01722} {\bibfield  {journal}
  {\bibinfo  {journal} {Nano Letters}\ }\textbf {\bibinfo {volume} {22}},\
  \bibinfo {pages} {6553} (\bibinfo {year} {2022})}\BibitemShut {NoStop}%
\bibitem [{\citenamefont {Liu}\ \emph {et~al.}(2021)\citenamefont {Liu},
  \citenamefont {Li}, \citenamefont {Yang}, \citenamefont {Yu}, \citenamefont
  {Meng}, \citenamefont {Wang}, \citenamefont {Li}, \citenamefont {Guo},
  \citenamefont {Yan}, \citenamefont {Li}, \citenamefont {Wang}, \citenamefont
  {Xu}, \citenamefont {Wang}, \citenamefont {Tang}, \citenamefont {Li},\ and\
  \citenamefont {Guo}}]{ACSPhot_Guo2021}%
  \BibitemOpen
  \bibfield  {author} {\bibinfo {author} {\bibfnamefont {W.}~\bibnamefont
  {Liu}}, \bibinfo {author} {\bibfnamefont {Z.-P.}\ \bibnamefont {Li}},
  \bibinfo {author} {\bibfnamefont {Y.-Z.}\ \bibnamefont {Yang}}, \bibinfo
  {author} {\bibfnamefont {S.}~\bibnamefont {Yu}}, \bibinfo {author}
  {\bibfnamefont {Y.}~\bibnamefont {Meng}}, \bibinfo {author} {\bibfnamefont
  {Z.-A.}\ \bibnamefont {Wang}}, \bibinfo {author} {\bibfnamefont {Z.-C.}\
  \bibnamefont {Li}}, \bibinfo {author} {\bibfnamefont {N.-J.}\ \bibnamefont
  {Guo}}, \bibinfo {author} {\bibfnamefont {F.-F.}\ \bibnamefont {Yan}},
  \bibinfo {author} {\bibfnamefont {Q.}~\bibnamefont {Li}}, \bibinfo {author}
  {\bibfnamefont {J.-F.}\ \bibnamefont {Wang}}, \bibinfo {author}
  {\bibfnamefont {J.-S.}\ \bibnamefont {Xu}}, \bibinfo {author} {\bibfnamefont
  {Y.-T.}\ \bibnamefont {Wang}}, \bibinfo {author} {\bibfnamefont {J.-S.}\
  \bibnamefont {Tang}}, \bibinfo {author} {\bibfnamefont {C.-F.}\ \bibnamefont
  {Li}}, \ and\ \bibinfo {author} {\bibfnamefont {G.-C.}\ \bibnamefont {Guo}},\
  }\href {\doibase 10.1021/acsphotonics.1c00320} {\bibfield  {journal}
  {\bibinfo  {journal} {ACS Photonics}\ }\textbf {\bibinfo {volume} {8}},\
  \bibinfo {pages} {1889} (\bibinfo {year} {2021})}\BibitemShut {NoStop}%
\bibitem [{\citenamefont {Qian}\ \emph {et~al.}(2022)\citenamefont {Qian},
  \citenamefont {Villafañe}, \citenamefont {Schalk}, \citenamefont {Astakhov},
  \citenamefont {Kentsch}, \citenamefont {Helm}, \citenamefont {Soubelet},
  \citenamefont {Wilson}, \citenamefont {Rizzato}, \citenamefont {Mohr},
  \citenamefont {Holleitner}, \citenamefont {Bucher}, \citenamefont {Stier},\
  and\ \citenamefont {Finley}}]{Qian2022}%
  \BibitemOpen
  \bibfield  {author} {\bibinfo {author} {\bibfnamefont {C.}~\bibnamefont
  {Qian}}, \bibinfo {author} {\bibfnamefont {V.}~\bibnamefont {Villafañe}},
  \bibinfo {author} {\bibfnamefont {M.}~\bibnamefont {Schalk}}, \bibinfo
  {author} {\bibfnamefont {G.~V.}\ \bibnamefont {Astakhov}}, \bibinfo {author}
  {\bibfnamefont {U.}~\bibnamefont {Kentsch}}, \bibinfo {author} {\bibfnamefont
  {M.}~\bibnamefont {Helm}}, \bibinfo {author} {\bibfnamefont {P.}~\bibnamefont
  {Soubelet}}, \bibinfo {author} {\bibfnamefont {N.~P.}\ \bibnamefont
  {Wilson}}, \bibinfo {author} {\bibfnamefont {R.}~\bibnamefont {Rizzato}},
  \bibinfo {author} {\bibfnamefont {S.}~\bibnamefont {Mohr}}, \bibinfo {author}
  {\bibfnamefont {A.~W.}\ \bibnamefont {Holleitner}}, \bibinfo {author}
  {\bibfnamefont {D.~B.}\ \bibnamefont {Bucher}}, \bibinfo {author}
  {\bibfnamefont {A.~V.}\ \bibnamefont {Stier}}, \ and\ \bibinfo {author}
  {\bibfnamefont {J.~J.}\ \bibnamefont {Finley}},\ }\href
  {https://doi.org/10.1021/acs.nanolett.2c00739} {\bibfield  {journal}
  {\bibinfo  {journal} {Nano Letters}\ }\textbf {\bibinfo {volume} {22}},\
  \bibinfo {pages} {5137} (\bibinfo {year} {2022})}\BibitemShut {NoStop}%
\bibitem [{\citenamefont {Libbi}\ \emph {et~al.}(2022)\citenamefont {Libbi},
  \citenamefont {de~Melo}, \citenamefont {Zanolli}, \citenamefont
  {Verstraete},\ and\ \citenamefont {Marzari}}]{Libbi_2022}%
  \BibitemOpen
  \bibfield  {author} {\bibinfo {author} {\bibfnamefont {F.}~\bibnamefont
  {Libbi}}, \bibinfo {author} {\bibfnamefont {P.~M. M.~C.}\ \bibnamefont
  {de~Melo}}, \bibinfo {author} {\bibfnamefont {Z.}~\bibnamefont {Zanolli}},
  \bibinfo {author} {\bibfnamefont {M.~J.}\ \bibnamefont {Verstraete}}, \ and\
  \bibinfo {author} {\bibfnamefont {N.}~\bibnamefont {Marzari}},\ }\href
  {\doibase 10.1103/PhysRevLett.128.167401} {\bibfield  {journal} {\bibinfo
  {journal} {Phys. Rev. Lett.}\ }\textbf {\bibinfo {volume} {128}},\ \bibinfo
  {pages} {167401} (\bibinfo {year} {2022})}\BibitemShut {NoStop}%
\bibitem [{\citenamefont {Gong}\ \emph {et~al.}(2022)\citenamefont {Gong},
  \citenamefont {He}, \citenamefont {Gao}, \citenamefont {Ju}, \citenamefont
  {Liu}, \citenamefont {Ye}, \citenamefont {Henriksen}, \citenamefont {Li},\
  and\ \citenamefont {Zu}}]{Gong_2022}%
  \BibitemOpen
  \bibfield  {author} {\bibinfo {author} {\bibfnamefont {R.}~\bibnamefont
  {Gong}}, \bibinfo {author} {\bibfnamefont {G.}~\bibnamefont {He}}, \bibinfo
  {author} {\bibfnamefont {X.}~\bibnamefont {Gao}}, \bibinfo {author}
  {\bibfnamefont {P.}~\bibnamefont {Ju}}, \bibinfo {author} {\bibfnamefont
  {Z.}~\bibnamefont {Liu}}, \bibinfo {author} {\bibfnamefont {B.}~\bibnamefont
  {Ye}}, \bibinfo {author} {\bibfnamefont {E.~A.}\ \bibnamefont {Henriksen}},
  \bibinfo {author} {\bibfnamefont {T.}~\bibnamefont {Li}}, \ and\ \bibinfo
  {author} {\bibfnamefont {C.}~\bibnamefont {Zu}},\ }\href@noop {} {\enquote
  {\bibinfo {title} {Coherent dynamics of strongly interacting electronic spin
  defects in hexagonal boron nitride},}\ } (\bibinfo {year} {2022}),\ \Eprint
  {http://arxiv.org/abs/2210.11485} {arXiv:2210.11485 [quant-ph]} \BibitemShut
  {NoStop}%
\bibitem [{\citenamefont {Barson}\ \emph {et~al.}(2017)\citenamefont {Barson},
  \citenamefont {Peddibhotla}, \citenamefont {Ovartchaiyapong}, \citenamefont
  {Ganesan}, \citenamefont {Taylor}, \citenamefont {Gebert}, \citenamefont
  {Mielens}, \citenamefont {Koslowski}, \citenamefont {Simpson}, \citenamefont
  {McGuinness}, \citenamefont {McCallum}, \citenamefont {Prawer}, \citenamefont
  {Onoda}, \citenamefont {Ohshima}, \citenamefont {Bleszynski~Jayich},
  \citenamefont {Jelezko}, \citenamefont {Manson},\ and\ \citenamefont
  {Doherty}}]{Barson_2017}%
  \BibitemOpen
  \bibfield  {author} {\bibinfo {author} {\bibfnamefont {M.~S.~J.}\
  \bibnamefont {Barson}}, \bibinfo {author} {\bibfnamefont {P.}~\bibnamefont
  {Peddibhotla}}, \bibinfo {author} {\bibfnamefont {P.}~\bibnamefont
  {Ovartchaiyapong}}, \bibinfo {author} {\bibfnamefont {K.}~\bibnamefont
  {Ganesan}}, \bibinfo {author} {\bibfnamefont {R.~L.}\ \bibnamefont {Taylor}},
  \bibinfo {author} {\bibfnamefont {M.}~\bibnamefont {Gebert}}, \bibinfo
  {author} {\bibfnamefont {Z.}~\bibnamefont {Mielens}}, \bibinfo {author}
  {\bibfnamefont {B.}~\bibnamefont {Koslowski}}, \bibinfo {author}
  {\bibfnamefont {D.~A.}\ \bibnamefont {Simpson}}, \bibinfo {author}
  {\bibfnamefont {L.~P.}\ \bibnamefont {McGuinness}}, \bibinfo {author}
  {\bibfnamefont {J.}~\bibnamefont {McCallum}}, \bibinfo {author}
  {\bibfnamefont {S.}~\bibnamefont {Prawer}}, \bibinfo {author} {\bibfnamefont
  {S.}~\bibnamefont {Onoda}}, \bibinfo {author} {\bibfnamefont
  {T.}~\bibnamefont {Ohshima}}, \bibinfo {author} {\bibfnamefont {A.~C.}\
  \bibnamefont {Bleszynski~Jayich}}, \bibinfo {author} {\bibfnamefont
  {F.}~\bibnamefont {Jelezko}}, \bibinfo {author} {\bibfnamefont {N.~B.}\
  \bibnamefont {Manson}}, \ and\ \bibinfo {author} {\bibfnamefont {M.~W.}\
  \bibnamefont {Doherty}},\ }\href
  {https://doi.org/10.1021/acs.nanolett.6b04544} {\bibfield  {journal}
  {\bibinfo  {journal} {Nano Letters}\ }\textbf {\bibinfo {volume} {17}},\
  \bibinfo {pages} {1496} (\bibinfo {year} {2017})}\BibitemShut {NoStop}%
\bibitem [{\citenamefont {{Van Oort}}\ and\ \citenamefont
  {Glasbeek}(1990)}]{VanOort_1990}%
  \BibitemOpen
  \bibfield  {author} {\bibinfo {author} {\bibfnamefont {E.}~\bibnamefont {{Van
  Oort}}}\ and\ \bibinfo {author} {\bibfnamefont {M.}~\bibnamefont
  {Glasbeek}},\ }\href {\doibase https://doi.org/10.1016/0009-2614(90)85665-Y}
  {\bibfield  {journal} {\bibinfo  {journal} {Chemical Physics Letters}\
  }\textbf {\bibinfo {volume} {168}},\ \bibinfo {pages} {529} (\bibinfo {year}
  {1990})}\BibitemShut {NoStop}%
\bibitem [{\citenamefont {Sajid}\ \emph
  {et~al.}(2020{\natexlab{c}})\citenamefont {Sajid}, \citenamefont {Thygesen},
  \citenamefont {Reimers},\ and\ \citenamefont {Ford}}]{Sajid_2020_edge}%
  \BibitemOpen
  \bibfield  {author} {\bibinfo {author} {\bibfnamefont {A.}~\bibnamefont
  {Sajid}}, \bibinfo {author} {\bibfnamefont {K.~S.}\ \bibnamefont {Thygesen}},
  \bibinfo {author} {\bibfnamefont {J.~R.}\ \bibnamefont {Reimers}}, \ and\
  \bibinfo {author} {\bibfnamefont {M.~J.}\ \bibnamefont {Ford}},\ }\href
  {\doibase 10.1038/s42005-020-00416-z} {\bibfield  {journal} {\bibinfo
  {journal} {Communications Physics}\ }\textbf {\bibinfo {volume} {3}},\
  \bibinfo {pages} {153} (\bibinfo {year} {2020}{\natexlab{c}})}\BibitemShut
  {NoStop}%
\bibitem [{\citenamefont {Gracheva}\ \emph {et~al.}(2023)\citenamefont
  {Gracheva}, \citenamefont {Murzakhanov}, \citenamefont {Mamin}, \citenamefont
  {Sadovnikova}, \citenamefont {Gabbasov}, \citenamefont {Mokhov},\ and\
  \citenamefont {Gafurov}}]{Gracheva2023}%
  \BibitemOpen
  \bibfield  {author} {\bibinfo {author} {\bibfnamefont {I.~N.}\ \bibnamefont
  {Gracheva}}, \bibinfo {author} {\bibfnamefont {F.~F.}\ \bibnamefont
  {Murzakhanov}}, \bibinfo {author} {\bibfnamefont {G.~V.}\ \bibnamefont
  {Mamin}}, \bibinfo {author} {\bibfnamefont {M.~A.}\ \bibnamefont
  {Sadovnikova}}, \bibinfo {author} {\bibfnamefont {B.~F.}\ \bibnamefont
  {Gabbasov}}, \bibinfo {author} {\bibfnamefont {E.~N.}\ \bibnamefont
  {Mokhov}}, \ and\ \bibinfo {author} {\bibfnamefont {M.~R.}\ \bibnamefont
  {Gafurov}},\ }\href {https://doi.org/10.1021/acs.jpcc.2c08716} {\bibfield
  {journal} {\bibinfo  {journal} {The Journal of Physical Chemistry C}\
  }\textbf {\bibinfo {volume} {127}},\ \bibinfo {pages} {3634} (\bibinfo {year}
  {2023})}\BibitemShut {NoStop}%
\bibitem [{\citenamefont {Udvarhelyi}\ \emph {et~al.}(2018)\citenamefont
  {Udvarhelyi}, \citenamefont {Shkolnikov}, \citenamefont {Gali}, \citenamefont
  {Burkard},\ and\ \citenamefont {P\'alyi}}]{Udvarhelyi_2018}%
  \BibitemOpen
  \bibfield  {author} {\bibinfo {author} {\bibfnamefont {P.}~\bibnamefont
  {Udvarhelyi}}, \bibinfo {author} {\bibfnamefont {V.~O.}\ \bibnamefont
  {Shkolnikov}}, \bibinfo {author} {\bibfnamefont {A.}~\bibnamefont {Gali}},
  \bibinfo {author} {\bibfnamefont {G.}~\bibnamefont {Burkard}}, \ and\
  \bibinfo {author} {\bibfnamefont {A.}~\bibnamefont {P\'alyi}},\ }\href
  {\doibase 10.1103/PhysRevB.98.075201} {\bibfield  {journal} {\bibinfo
  {journal} {Phys. Rev. B}\ }\textbf {\bibinfo {volume} {98}},\ \bibinfo
  {pages} {075201} (\bibinfo {year} {2018})}\BibitemShut {NoStop}%
\bibitem [{\citenamefont {Bosak}\ \emph {et~al.}(2006)\citenamefont {Bosak},
  \citenamefont {Serrano}, \citenamefont {Krisch}, \citenamefont {Watanabe},
  \citenamefont {Taniguchi},\ and\ \citenamefont {Kanda}}]{Bosak_2006}%
  \BibitemOpen
  \bibfield  {author} {\bibinfo {author} {\bibfnamefont {A.}~\bibnamefont
  {Bosak}}, \bibinfo {author} {\bibfnamefont {J.}~\bibnamefont {Serrano}},
  \bibinfo {author} {\bibfnamefont {M.}~\bibnamefont {Krisch}}, \bibinfo
  {author} {\bibfnamefont {K.}~\bibnamefont {Watanabe}}, \bibinfo {author}
  {\bibfnamefont {T.}~\bibnamefont {Taniguchi}}, \ and\ \bibinfo {author}
  {\bibfnamefont {H.}~\bibnamefont {Kanda}},\ }\href {\doibase
  10.1103/PhysRevB.73.041402} {\bibfield  {journal} {\bibinfo  {journal} {Phys.
  Rev. B}\ }\textbf {\bibinfo {volume} {73}},\ \bibinfo {pages} {041402}
  (\bibinfo {year} {2006})}\BibitemShut {NoStop}%
\bibitem [{\citenamefont {Mittiga}\ \emph {et~al.}(2018)\citenamefont
  {Mittiga}, \citenamefont {Hsieh}, \citenamefont {Zu}, \citenamefont {Kobrin},
  \citenamefont {Machado}, \citenamefont {Bhattacharyya}, \citenamefont {Rui},
  \citenamefont {Jarmola}, \citenamefont {Choi}, \citenamefont {Budker},\ and\
  \citenamefont {Yao}}]{Mittiga_2018}%
  \BibitemOpen
  \bibfield  {author} {\bibinfo {author} {\bibfnamefont {T.}~\bibnamefont
  {Mittiga}}, \bibinfo {author} {\bibfnamefont {S.}~\bibnamefont {Hsieh}},
  \bibinfo {author} {\bibfnamefont {C.}~\bibnamefont {Zu}}, \bibinfo {author}
  {\bibfnamefont {B.}~\bibnamefont {Kobrin}}, \bibinfo {author} {\bibfnamefont
  {F.}~\bibnamefont {Machado}}, \bibinfo {author} {\bibfnamefont
  {P.}~\bibnamefont {Bhattacharyya}}, \bibinfo {author} {\bibfnamefont {N.~Z.}\
  \bibnamefont {Rui}}, \bibinfo {author} {\bibfnamefont {A.}~\bibnamefont
  {Jarmola}}, \bibinfo {author} {\bibfnamefont {S.}~\bibnamefont {Choi}},
  \bibinfo {author} {\bibfnamefont {D.}~\bibnamefont {Budker}}, \ and\ \bibinfo
  {author} {\bibfnamefont {N.~Y.}\ \bibnamefont {Yao}},\ }\href {\doibase
  10.1103/PhysRevLett.121.246402} {\bibfield  {journal} {\bibinfo  {journal}
  {Phys. Rev. Lett.}\ }\textbf {\bibinfo {volume} {121}},\ \bibinfo {pages}
  {246402} (\bibinfo {year} {2018})}\BibitemShut {NoStop}%
\bibitem [{\citenamefont {Liu}\ \emph {et~al.}(2022)\citenamefont {Liu},
  \citenamefont {Iv{\'a}dy}, \citenamefont {Li}, \citenamefont {Yang},
  \citenamefont {Yu}, \citenamefont {Meng}, \citenamefont {Wang}, \citenamefont
  {Guo}, \citenamefont {Yan}, \citenamefont {Li}, \citenamefont {Wang},
  \citenamefont {Xu}, \citenamefont {Liu}, \citenamefont {Zhou}, \citenamefont
  {Dong}, \citenamefont {Chen}, \citenamefont {Sun}, \citenamefont {Wang},
  \citenamefont {Tang}, \citenamefont {Gali}, \citenamefont {Li},\ and\
  \citenamefont {Guo}}]{Liu2022}%
  \BibitemOpen
  \bibfield  {author} {\bibinfo {author} {\bibfnamefont {W.}~\bibnamefont
  {Liu}}, \bibinfo {author} {\bibfnamefont {V.}~\bibnamefont {Iv{\'a}dy}},
  \bibinfo {author} {\bibfnamefont {Z.-P.}\ \bibnamefont {Li}}, \bibinfo
  {author} {\bibfnamefont {Y.-Z.}\ \bibnamefont {Yang}}, \bibinfo {author}
  {\bibfnamefont {S.}~\bibnamefont {Yu}}, \bibinfo {author} {\bibfnamefont
  {Y.}~\bibnamefont {Meng}}, \bibinfo {author} {\bibfnamefont {Z.-A.}\
  \bibnamefont {Wang}}, \bibinfo {author} {\bibfnamefont {N.-J.}\ \bibnamefont
  {Guo}}, \bibinfo {author} {\bibfnamefont {F.-F.}\ \bibnamefont {Yan}},
  \bibinfo {author} {\bibfnamefont {Q.}~\bibnamefont {Li}}, \bibinfo {author}
  {\bibfnamefont {J.-F.}\ \bibnamefont {Wang}}, \bibinfo {author}
  {\bibfnamefont {J.-S.}\ \bibnamefont {Xu}}, \bibinfo {author} {\bibfnamefont
  {X.}~\bibnamefont {Liu}}, \bibinfo {author} {\bibfnamefont {Z.-Q.}\
  \bibnamefont {Zhou}}, \bibinfo {author} {\bibfnamefont {Y.}~\bibnamefont
  {Dong}}, \bibinfo {author} {\bibfnamefont {X.-D.}\ \bibnamefont {Chen}},
  \bibinfo {author} {\bibfnamefont {F.-W.}\ \bibnamefont {Sun}}, \bibinfo
  {author} {\bibfnamefont {Y.-T.}\ \bibnamefont {Wang}}, \bibinfo {author}
  {\bibfnamefont {J.-S.}\ \bibnamefont {Tang}}, \bibinfo {author}
  {\bibfnamefont {A.}~\bibnamefont {Gali}}, \bibinfo {author} {\bibfnamefont
  {C.-F.}\ \bibnamefont {Li}}, \ and\ \bibinfo {author} {\bibfnamefont {G.-C.}\
  \bibnamefont {Guo}},\ }\href {\doibase 10.1038/s41467-022-33399-2} {\bibfield
   {journal} {\bibinfo  {journal} {Nature Communications}\ }\textbf {\bibinfo
  {volume} {13}},\ \bibinfo {pages} {5713} (\bibinfo {year}
  {2022})}\BibitemShut {NoStop}%
\bibitem [{\citenamefont {Falk}\ \emph {et~al.}(2014)\citenamefont {Falk},
  \citenamefont {Klimov}, \citenamefont {Buckley}, \citenamefont {Iv\'ady},
  \citenamefont {Abrikosov}, \citenamefont {Calusine}, \citenamefont {Koehl},
  \citenamefont {Gali},\ and\ \citenamefont {Awschalom}}]{Falk_2014}%
  \BibitemOpen
  \bibfield  {author} {\bibinfo {author} {\bibfnamefont {A.~L.}\ \bibnamefont
  {Falk}}, \bibinfo {author} {\bibfnamefont {P.~V.}\ \bibnamefont {Klimov}},
  \bibinfo {author} {\bibfnamefont {B.~B.}\ \bibnamefont {Buckley}}, \bibinfo
  {author} {\bibfnamefont {V.}~\bibnamefont {Iv\'ady}}, \bibinfo {author}
  {\bibfnamefont {I.~A.}\ \bibnamefont {Abrikosov}}, \bibinfo {author}
  {\bibfnamefont {G.}~\bibnamefont {Calusine}}, \bibinfo {author}
  {\bibfnamefont {W.~F.}\ \bibnamefont {Koehl}}, \bibinfo {author}
  {\bibfnamefont {A.}~\bibnamefont {Gali}}, \ and\ \bibinfo {author}
  {\bibfnamefont {D.~D.}\ \bibnamefont {Awschalom}},\ }\href {\doibase
  10.1103/PhysRevLett.112.187601} {\bibfield  {journal} {\bibinfo  {journal}
  {Phys. Rev. Lett.}\ }\textbf {\bibinfo {volume} {112}},\ \bibinfo {pages}
  {187601} (\bibinfo {year} {2014})}\BibitemShut {NoStop}%
\bibitem [{\citenamefont {Adamo}\ and\ \citenamefont {Barone}(1999)}]{PBE0}%
  \BibitemOpen
  \bibfield  {author} {\bibinfo {author} {\bibfnamefont {C.}~\bibnamefont
  {Adamo}}\ and\ \bibinfo {author} {\bibfnamefont {V.}~\bibnamefont {Barone}},\
  }\href {\doibase 10.1063/1.478522} {\bibfield  {journal} {\bibinfo  {journal}
  {The Journal of Chemical Physics}\ }\textbf {\bibinfo {volume} {110}},\
  \bibinfo {pages} {6158} (\bibinfo {year} {1999})}\BibitemShut {NoStop}%
\bibitem [{\citenamefont {Weigend}\ and\ \citenamefont
  {Ahlrichs}(2005)}]{def2}%
  \BibitemOpen
  \bibfield  {author} {\bibinfo {author} {\bibfnamefont {F.}~\bibnamefont
  {Weigend}}\ and\ \bibinfo {author} {\bibfnamefont {R.}~\bibnamefont
  {Ahlrichs}},\ }\href {\doibase 10.1039/B508541A} {\bibfield  {journal}
  {\bibinfo  {journal} {Phys. Chem. Chem. Phys.}\ }\textbf {\bibinfo {volume}
  {7}},\ \bibinfo {pages} {3297} (\bibinfo {year} {2005})}\BibitemShut
  {NoStop}%
\bibitem [{\citenamefont {Neese}(2012)}]{ORCA}%
  \BibitemOpen
  \bibfield  {author} {\bibinfo {author} {\bibfnamefont {F.}~\bibnamefont
  {Neese}},\ }\href {\doibase 10.1002/wcms.81} {\bibfield  {journal} {\bibinfo
  {journal} {WIREs Computational Molecular Science}\ }\textbf {\bibinfo
  {volume} {2}},\ \bibinfo {pages} {73} (\bibinfo {year} {2012})}\BibitemShut
  {NoStop}%
\bibitem [{\citenamefont {Sinnecker}\ and\ \citenamefont
  {Neese}(2006)}]{Sinnecker_2006}%
  \BibitemOpen
  \bibfield  {author} {\bibinfo {author} {\bibfnamefont {S.}~\bibnamefont
  {Sinnecker}}\ and\ \bibinfo {author} {\bibfnamefont {F.}~\bibnamefont
  {Neese}},\ }\href {https://doi.org/10.1021/jp0643303} {\bibfield  {journal}
  {\bibinfo  {journal} {The Journal of Physical Chemistry A}\ }\textbf
  {\bibinfo {volume} {110}},\ \bibinfo {pages} {12267} (\bibinfo {year}
  {2006})}\BibitemShut {NoStop}%
\bibitem [{\citenamefont {Liu}\ \emph {et~al.}(2018)\citenamefont {Liu},
  \citenamefont {He}, \citenamefont {Xue}, \citenamefont {Li}, \citenamefont
  {Liu},\ and\ \citenamefont {Edgar}}]{Liu2018}%
  \BibitemOpen
  \bibfield  {author} {\bibinfo {author} {\bibfnamefont {S.}~\bibnamefont
  {Liu}}, \bibinfo {author} {\bibfnamefont {R.}~\bibnamefont {He}}, \bibinfo
  {author} {\bibfnamefont {L.}~\bibnamefont {Xue}}, \bibinfo {author}
  {\bibfnamefont {J.}~\bibnamefont {Li}}, \bibinfo {author} {\bibfnamefont
  {B.}~\bibnamefont {Liu}}, \ and\ \bibinfo {author} {\bibfnamefont {J.~H.}\
  \bibnamefont {Edgar}},\ }\href
  {https://doi.org/10.1021/acs.chemmater.8b02589} {\bibfield  {journal}
  {\bibinfo  {journal} {Chem. Mater.}\ }\textbf {\bibinfo {volume} {30}},\
  \bibinfo {pages} {6222} (\bibinfo {year} {2018})}\BibitemShut {NoStop}%
\end{thebibliography}%
\bibliographystyle{apsrev4-1}

\end{document}